\documentclass[iop]{emulateapj}

\usepackage{rotating}
\usepackage{xcolor}

\def\s{\phantom}

\def\amin{\ifmmode^{\prime}\else$^{\prime}$\fi}
\def\asec{\ifmmode^{\prime\prime}\else$^{\prime\prime}$\fi}

\def\simgt{\lower.5ex\hbox{$\; \buildrel > \over \sim \;$}}
\def\simlt{\lower.5ex\hbox{$\; \buildrel < \over \sim \;$}}

\newcommand\chandra{{\it Chandra}}

\newcommand\xmm{{\it XMM-Newton}}

\newcommand\integral{{\it INTEGRAL}/IBIS}

\newcommand\swift{{\it Swift\/}}
\newcommand\nustar{\hbox{\it NuSTAR\/}}

\newcommand\eflux{erg\,cm$^{-2}$\,s$^{-1}$}

\newcommand\axj{AX J1745.6$-$2901}

\graphicspath{{./}{figures/}}

\shorttitle{X-ray transients in the Galactic Center}
\shortauthors{Mori et al.}
\begin{document}

\title{{\it NuSTAR} and {\it Chandra} observations of new X-ray transients in the central parsec of the Galaxy}

\author{Kaya Mori\altaffilmark{1}, Charles~J.~Hailey\altaffilmark{1}, Shifra~Mandel\altaffilmark{1}, Yve~E.~Schutt\altaffilmark{1}, Matteo~Bachetti\altaffilmark{2,3}, Anna~Coerver\altaffilmark{1}, Frederick~K.~Baganoff\altaffilmark{4}, Hannah~Dykaar\altaffilmark{5}, Jonathan~E.~Grindlay\altaffilmark{6}, Daryl~Haggard\altaffilmark{7},
Keri~Heuer\altaffilmark{8},
Jaesub~Hong\altaffilmark{6}, 
Benjamin~J.~Hord\altaffilmark{9},
Chichuan~Jin\altaffilmark{10}, Melania~Nynka\altaffilmark{4}, Gabriele~Ponti\altaffilmark{11}, John~A.~Tomsick\altaffilmark{12}}

%\author[0000-0002-0786-7307]{Kaya Mori}
%\affil{Columbia Astrophysics Laboratory 550 West 120th Street, New York, NY 10023, USA}
\altaffiltext{1}{Columbia Astrophysics Laboratory, Columbia University, New York, NY 10027, USA}
\altaffiltext{2}{INAF-Osservatorio Astronomico di Cagliari, via della Scienza 5, I-09047 Selargius, Italy}
\altaffiltext{3}{Space Radiation Laboratory, Caltech, 1200 E California Blvd, Pasadena, CA 91125}
\altaffiltext{4}{Kavli Institute for Astrophysics and Space Research, Massachusetts Institute of Technology, Cambridge, MA 02139, USA}
\altaffiltext{5}{Department of Astronomy and Astrophysics, University of Toronto, 50 St George St, Toronto, ON M5S 3H4, Canada} 
\altaffiltext{6}{Harvard-Smithsonian Center for Astrophysics, Cambridge, MA 02138, USA}
\altaffiltext{7}{McGill Space Institute and Department of Physics, McGill University, 3600 rue University, Montreal, Quebec, H3A 2T8, Canada}
\altaffiltext{8}{Department of Physics, Cornell University 109 Clark Hall, Ithaca, New York 14853-2501, USA}
\altaffiltext{9}{Department of Astronomy, University of Maryland College Park, MD 20742-2421, USA}
\altaffiltext{10}{National Astronomical Observatories (NAOC), CAS A20, Datun Road, Chaoyang District 100101, Beijing, P. R. China}
\altaffiltext{11}{INAF - Osservatorio Astronomico di Brera, via E. Bianchi 46, I-23807 Merate, Italy}
\altaffiltext{12}{Space Sciences Laboratory, University of California, Berkeley, CA 94720, USA}

\email{kaya@astro.columbia.edu}

\begin{abstract}

We report \nustar\ and \chandra\ observations of two X-ray transients, SWIFT~J174540.7$-$290015 (T15) and SWIFT~J174540.2$-$290037 (T37), which were discovered by the {\it Neil Gehrels Swift Observatory} in 2016 within $r\sim1$ pc of Sgr A*. \nustar\ detected bright X-ray outbursts from T15 and T37, likely in the soft and hard states, with 3--79~keV luminosities of  $8\times10^{36}$ and  $3\times10^{37}$~erg\,s$^{-1}$, respectively.  
No X-ray outbursts have previously been detected from the two transients and our \chandra\ ACIS analysis puts an upper limit of $L_X \simlt 2 \times10^{31}$~erg\,s$^{-1}$ on their quiescent 2--8~keV luminosities.  
No pulsations, significant QPOs, or type I X-ray bursts were detected in the \nustar\ data. While T15 exhibited no significant red noise, the T37 power density spectra are well characterized by three Lorentzian components. The declining variability of T37 above $\nu \sim 10$~Hz is typical of black hole (BH) transients in the hard state. \nustar\ spectra of both transients exhibit a thermal disk blackbody, X-ray reflection with broadened Fe atomic features, and a continuum component well described by Comptonization models. Their X-ray reflection spectra are most consistent with high BH spin ($a_{*}\simgt0.9$) and large disk density ($n_e\sim10^{21}$~cm$^{-3}$). Based on the best-fit ionization parameters and disk densities, we found that X-ray reflection occurred near the inner disk radius, which was derived from the relativistic broadening and thermal disk component. 
These X-ray characteristics suggest the outbursting BH-LMXB scenario for both transients and yield the first BH spin measurements from X-ray transients in the central 100 parsec region.

\end{abstract}

\keywords{Galaxy: center ---- X-rays: binaries --- X-rays: bursts}


\section{Introduction} \label{sec:intro}

The recent discovery of a dozen quiescent X-ray  binaries (XRBs) within a parsec of Sgr A* \citep{Hailey2018} confirmed the fundamental prediction that a density cusp of compact objects exists near a supermassive BH \citep{Bahcall1976, Bahcall1977, Morris1993, Miralda2000}.
The properties of these XRBs and their luminosity function point to a large population of hundreds of LMXBs in the central parsec.
The high concentration of LMXBs is in contrast to the larger spatial extent of the magnetic cataclysmic variable (CV) population over the central 10 pc region \citep{Perez2015, Hailey2016, Hong2016, Zhu2018}.
Earlier X-ray observations also suggested an overabundance of X-ray transients, with occasional outbursts from black hole (BH) and neutron star (NS) LMXBs, lasting for weeks to months, in the central parsec \citep{Muno2005}. Detecting more X-ray transients and identifying XRBs in quiescence are important for testing some theoretical predictions of the BH/NS population and XRB formation in the GC \citep{Generozov2018, Szolgyen2018, Baumgardt2018, Panamarev2019}. 

Since 2006 February, daily \swift\ monitoring of a 25\amin$\times$25\amin\ region around Sgr A* (except when the GC is not visible from  November to February annually), has resulted in the detection of a dozen X-ray transients within $\sim20$~pc of the GC \citep{Degenaar2015}, including
a new transient magnetar \citep{Mori2013, Kennea2013}. Some of the X-ray transients have been identified as NS-LMXBs (e.g., \axj) with the detection of type I X-ray bursts \citep{Degenaar2012}. A subclass of X-ray transients called very faint X-ray transients (VFXTs), with peak X-ray luminosity below $10^{36}$~erg\,s$^{-1}$, was
also revealed by the \swift\ monitoring program, although its nature is not fully  understood.

In 2016, \swift\ detected two new X-ray transients, Swift~J174540.7$-$290015 (T15) and Swift~J174540.2$-$290037 (T37), within 1~pc from Sgr A*. \nustar\ performed Target of Opportunity (ToO)
observations to characterize these X-ray transients within $\simlt$2 weeks of the \swift\ detections. \nustar, with its broad (3--79~keV) energy band, $10 \mu$sec timing resolution, and minimal deadtime effects, is ideal for studying and identifying bright 
X-ray transients. With sub-arcminute angular resolution, \nustar\ was able to 
resolve the X-ray transients from other bright sources  in the GC, including the nearby outbursting NS-LMXB \axj\ \citep{Ponti2018}.

\begin{figure*}[ht!]
\plottwo{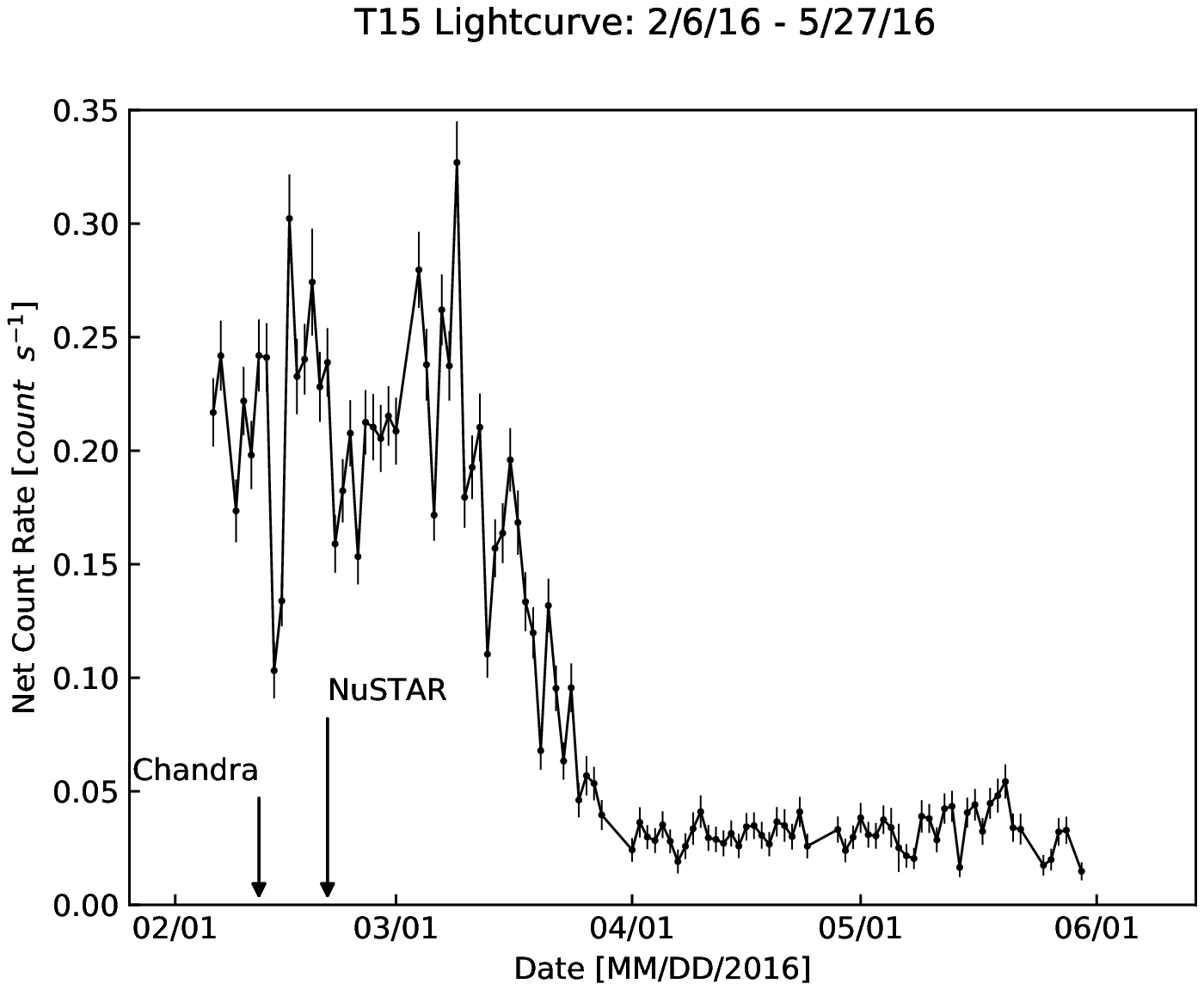}{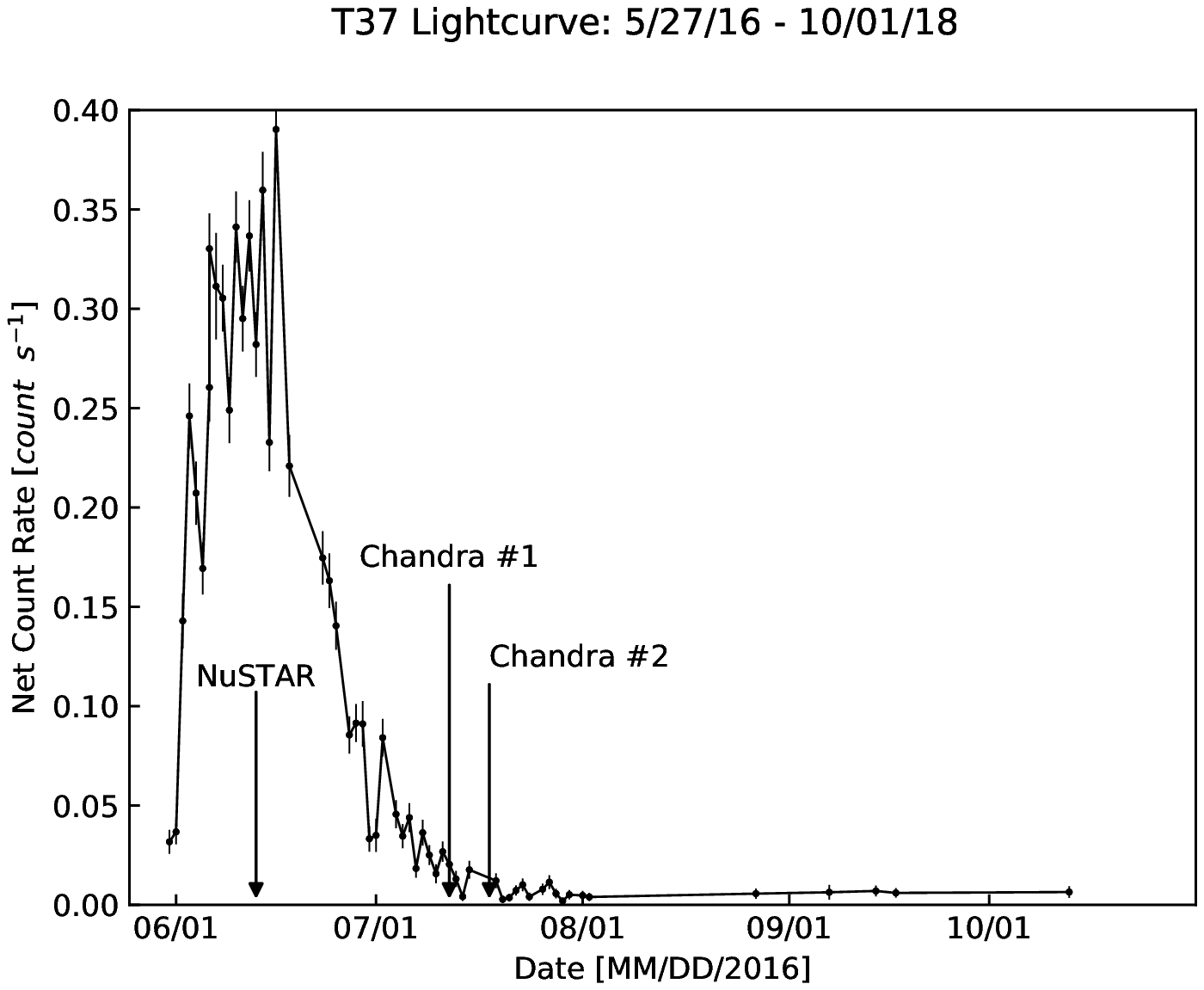}
\caption{\swift\ 2--10 keV lightcurves of T15 (left) and T37 (right) {  with 1-$\sigma$ statistical errors on the net counts}. For T15, \swift\ XRT net count rates were calculated by extracting source counts from a $r= 15$--40\asec\ annular region around the \chandra\ position and subtracting background counts from a source-free region of equivalent size.  For T37, a  similar annulus was used for extracting source photons but with the half closest to T15 removed from the region to avoid contamination from T15. The \nustar\ and \chandra\ observation dates are indicated by arrows. Note that the lightcurve of T15 is shown from 2016 February 6, when the \swift\ monitoring of the GC resumed, to 2016 May 27, when the T37 outburst began.
}
\label{fig:swift_lc}
\end{figure*}

This paper presents \nustar, \chandra, and \swift\ observations of the two X-ray transients that were conducted in 2016, and demonstrates how these follow-up X-ray observations can help us infer the nature of transient XRBs in the crowded GC region. We begin by reporting the X-ray observations of the two X-ray transients 
(\S\ref{sec:obs}). We describe \nustar\ spectral and timing analyses in \S\ref{sec:spectra} and \S\ref{sec:timing}, respectively. Then, in \S\ref{sec:chandra}, we present the \chandra\ data of the 2016 transients during the outbursts and in quiescence. Finally, we summarize our results and discuss the nature of the transients in \S\ref{sec:discussion}.  Throughout the paper, we assume a distance to the GC  of 8~kpc \citep{Reid1993, Camarillo2018}.

\section{X-ray observations and data reduction} \label{sec:obs}

On 2016 February 6, after \swift\ resumed the daily GC monitoring program following its hiatus due to the solar constraint window, an X-ray transient dubbed Swift~J174540.7$-$290015 (T15 hereafter) was discovered $\sim$16\asec\ north of Sgr A* \citep{Reynolds2016}. Follow-up \chandra\ observations on 2016 February 13--14 localized T15 at RA = 17:45:40.664$\pm$0.3433\asec\ and DEC = $-$29:00:15.61$\pm$0.3263\asec\ and confirmed it as a new X-ray transient \citep{Baganoff2016}. 
T15 was also observed by \xmm\ 
on 2016 February 26 and by \integral\ on 2016 February 11.  A detailed analysis of those observations, as well as GROND IR data and VLA radio observations, can be found in \citet{Ponti2016}.  All X-ray observations of the two \swift\ transients are summarized in Table~1. 
The exact start date of the T15 outburst is unknown, but it occurred sometime between 2015 November 2 and 2016 February 6 (when the GC was outside the \swift\ visibility window).

On 2016 May 28, while T15 was still in outburst, \swift\  discovered another new transient,  Swift~J174540.2$-$290037 (T37 hereafter), at RA = 17:45:40.60 and DEC = $-$29:00:36.4 (J2000) with an uncertainty of 3.5\asec\ (90\% C.L.), $\sim$10\asec\ south of Sgr A* \citep{Degenaar2016}. (Note that we determined the more accurate position of T37 using the \chandra\ observation data (see \S\ref{sec:chandra_obs})). T37 remained bright for about one month, during which the \nustar\ observation took place; subsequently, the X-ray flux rapidly decayed, as evidenced by two \chandra\ observations that were performed later. Neither of the X-ray transients has a counterpart in the \chandra\ X-ray source catalogs of \citet{Muno2009} and \citet{Zhu2018}.

\subsection{{\it Swift} observations and lightcurves}

We analyzed all \swift/XRT observations obtained in the Photon Counting (PC) mode from 2016 February 6 to 2016 October 1. Source photons of T15 and T37 were extracted using an $r=15$--40\asec\ annulus around the \chandra\ position to avoid pile-up. 
{  Note that the extraction region is much larger than the detector pixel size of 2.36\asec.} 
For T37, we excluded the annular half closest to T15 to avoid contamination. 
Background count rates were calculated from a nearby source-free region. {  As a result of excising a large part of the PSF, we ended up collecting $\sim20$\% and $\sim10$\% of the source photons for T15 and T37, 
respectively\footnote{In-flight calibration document on the \swift/XRT PSF available at \url{https://www.swift.ac.uk/analysis/xrt/files/Moretti\_spie\_xrtpsf2005.pdf}}. In addition, both the PSF and dust scattering halo profile are subject to large errors at $r \simgt 20$\asec. These systematic effects can lead to some uncertainty in the absolute X-ray flux measurements  based solely on \swift/XRT data. Indeed, we found that the \swift\ XRT fluxes were lower than those of \nustar\ by $\sim20$\% in the 3--10~keV band. Nevertheless, as \citet{Ponti2016} presented for T15, daily \swift/XRT data are useful for   studying the time evolution of the transients. Hence, we limited our usage of \swift/XRT data to constructing X-ray lightcurves. } 

Figure \ref{fig:swift_lc} shows 2--10 keV \swift/XRT net count rates of the two transients.  The T15 outburst lasted for at least  $\sim50$ days after its initial detection by \swift. Note that the duration of the T15 outburst could have been longer by up to 3 months, since the GC was not visible to \swift\ from the beginning of 2015 November. The \swift/XRT count rate of T15 stayed high at $\sim0.15$--0.3~ct\,s$^{-1}$ before it started decaying in mid 2016 March. T15 remained well above the background level until the T37 outburst began on 2016 May 28. On the other hand, the T37 lightcurve is characterized by a fast rise to the peak $\sim2$ weeks after the onset of the outburst and an exponential decay over $\sim$~30 days. \nustar\ observed T37 as the outburst was approaching its peak. The duration of the T37 outburst  ($\sim 30$ days) was shorter than that of T15 ($\simgt 50$ days).

%\startlongtable
\begin{deluxetable*}{c|c|c|c|c|l}
\tablecaption{Timeline of the two transients and X-ray observations in 2016 }
\tablehead{
\colhead{Date} & \colhead{ObsID} & \colhead{Target} & \colhead{Telescope} & \colhead{Exposure [ksec]} & \colhead{Comments} 
}
%\colnumbers
\startdata 
2016 Feb 6 & 00092201197 & T15 & \swift/XRT & ~1  &  
The first detection of the T15 outburst \\
2016 Feb 11 & 13200010001  & T15 & \integral\ & 10.8 & \\ 
2016 Feb 13 & 18055 & T15 & \chandra\  & 22.7 & \\
2016 Feb 14 & 18056 & T15 & \chandra\  & 21.8 & \\
2016 Feb 22 & 90101022002 & T15 & \nustar\ & 34 & \\ 
2016 Feb 26 & 0790180401 & T15 & \xmm\  & 35 & \\
2016 May 28 & 00092236057 & T37 & \swift/XRT & ~0.9 & The onset of the T37 outburst \\
2016 Jun 9  & 90201026002 & T37 & \nustar\  & 49 &  \\
2016 Jul 12 & 18731 & T37 & \chandra\ & 78.4 &  \\
2016 Jul 18 & 18732 & T37 & \chandra\ & 76.6 & 
\enddata
%\tablenotetext{a}{ }
%\tablecomments{Note}
\label{tab:obs} 
\end{deluxetable*}

\subsection{{\it NuSTAR} observations}

{\it NuSTAR} is composed of a pair of co-aligned high-energy X-ray focusing telescopes with focal plane modules FPMA and FPMB, which have an imaging resolution of 18\asec\ FWHM over a range of 3--79 keV and a characteristic 400 eV FWHM spectral resolution at 10 keV \citep{Harrison2013}. The absolute and relative timing accuracy of  \nustar, after correcting for on-board clock drift, are 3 msec and 10 $\mu$sec, respectively \citep{Madsen2015}. 

On 2016 February 22, 16 days after the first \swift\  detection of the T15 outburst, a 34 ks \nustar\ ToO observation was performed.
A 49 ks \nustar\ ToO observation of T37 was obtained on 2016 June 9, 11 days after the onset of the T37 outburst. The \nustar\ data of the two transients were reduced using NUSTARDAS v1.7.1. 

\begin{figure*}[ht!]
\plottwo{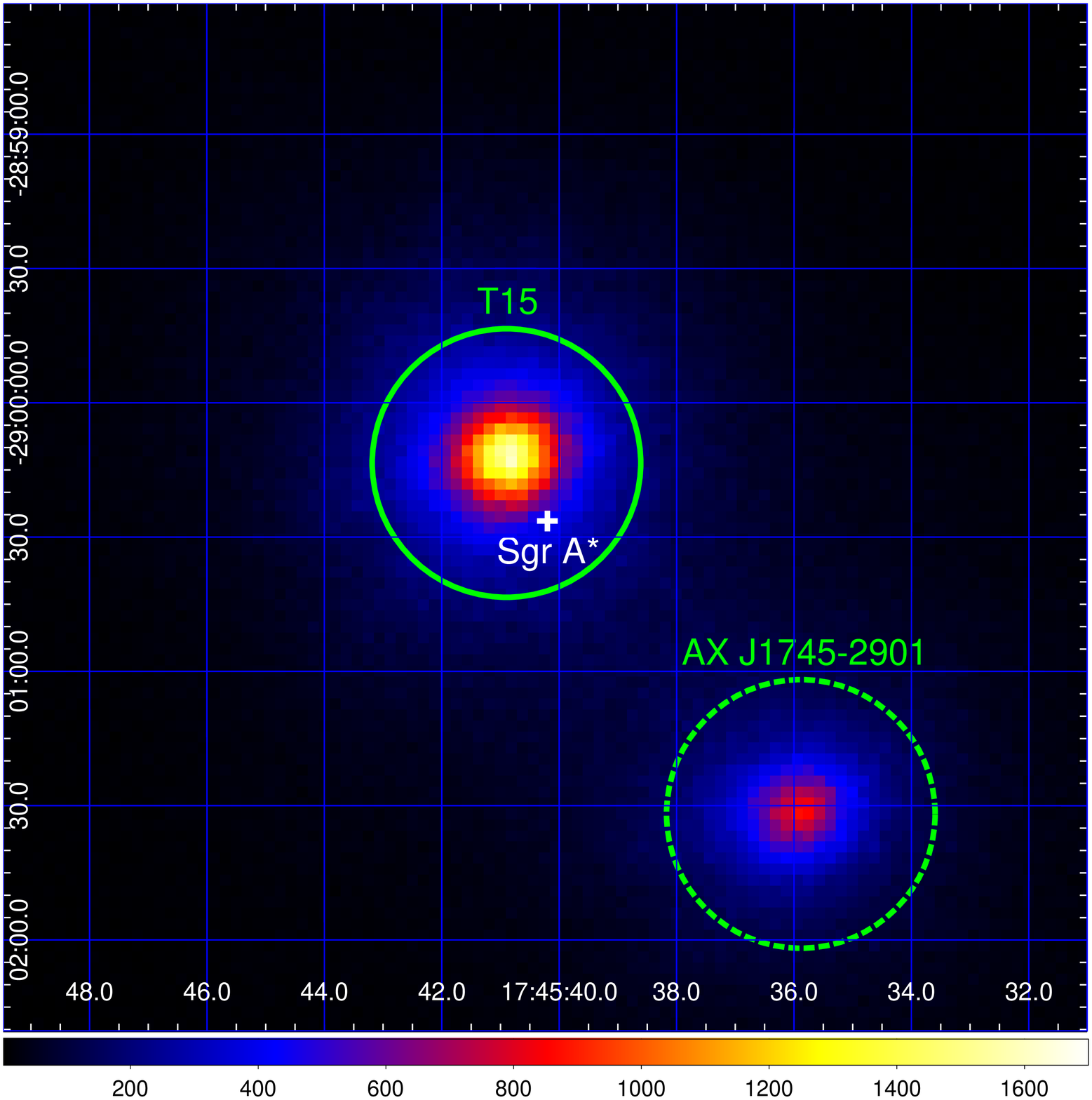}{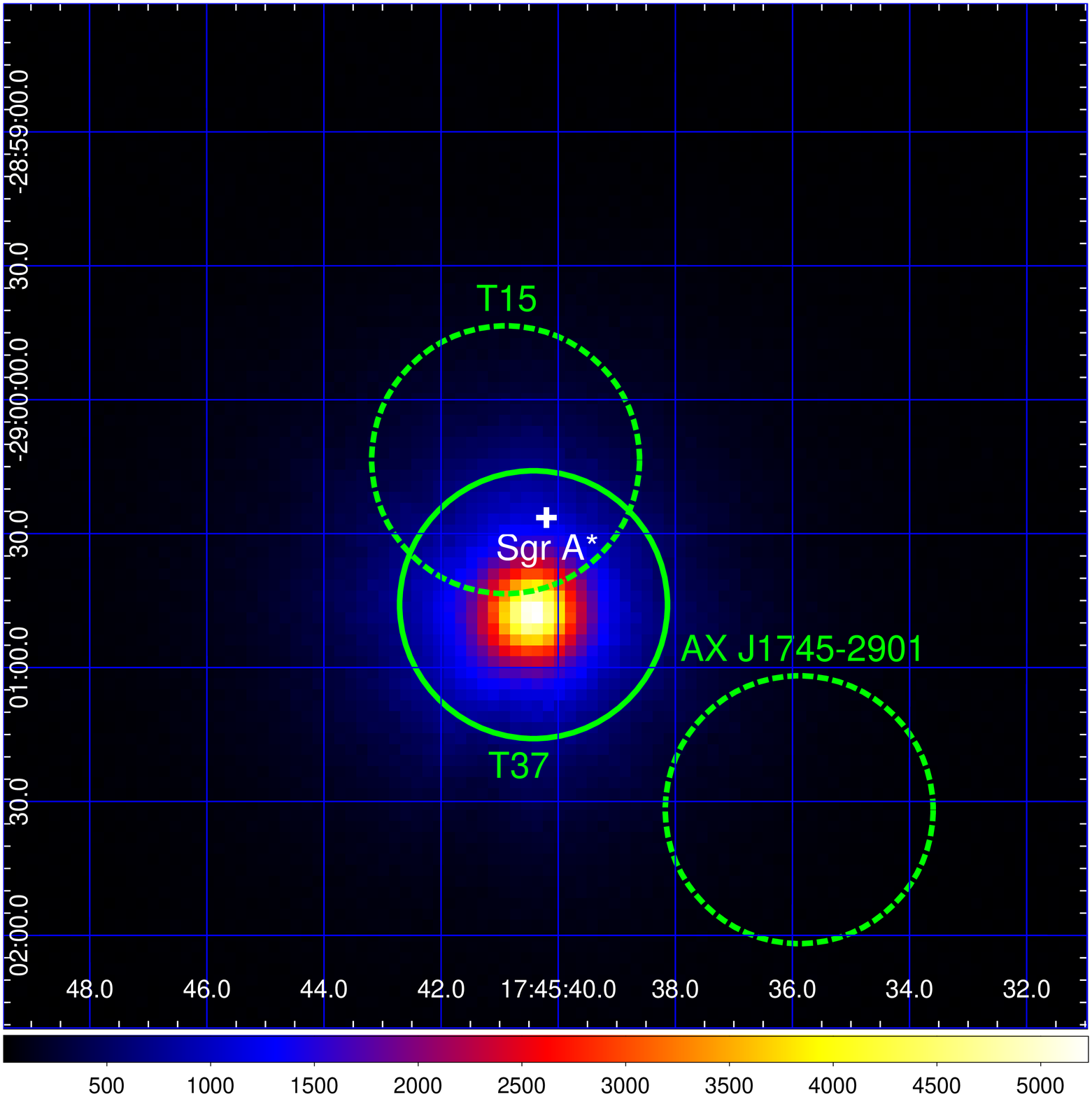}
\caption{{  \nustar\ FPMA 3--79 keV images from the  February (left) and June (right) 2016 observations. In the left image, solid and dashed circles (with a 30\asec\ radius) in green indicate T15 and \axj, respectively. In the right image,  a solid green circle (with a 30\asec\ radius) shows the location of T37 whose X-ray emission dominated over T15 and \axj\ (which are indicated in dashed green circles). The location of Sgr A* is indicated by a white cross near the center of the images.} 
}
\label{fig:nustar_images}
\end{figure*}

During the \nustar\ observations, emission from the two transients was dominant over background and other X-ray sources in the GC. {  Figure~\ref{fig:nustar_images} shows \nustar\ 3--79~keV images from the February and June 2016 observations. During the February 2016 observation, \nustar\ detected two transients, T15 and \axj. During the June 2016 observation, T37 was by far the brightest X-ray source in the GC, whereas X-ray emission from T15 and \axj\ had decayed significantly; thus  they are invisible in the \nustar\ images.} Source photons extracted from a $r=30$\asec\ circle (HPD) around the \chandra\ position give \nustar\ 3--79~keV count rates of 6.07/6.10 ct\,s$^{-1}$ for T15 and 13.2/12.6 (FPMA/FPMB) ct\,s$^{-1}$ for T37.

\subsection{{\it Chandra} observations}
\label{sec:chandra_obs}

\chandra\ observations  of T15 were performed on 2016 February 13 and 14 for 25 ks each, $\sim9$~days prior to the \nustar\ observation, with ACIS-S operating in the 1/8-subarray mode.  T37 was  observed, also in the 1/8-subarray mode, on 2016 July 12 and 18 for 78.4 and 76.6 ks, respectively, $\sim33$--39 days after the \nustar\ observation. 
The \chandra\ observations localized the two transients to better than 1\asec\ accuracy and the source radial profiles were used to determine the dust scattering parameters for T15 \citep{Corrales2017}. 
After registering the magnetar SGR~J1745$-$2900 to its  radio position, we determined the T37 position at RA = 17:45:40.42 and DEC = -29:00:45.93 (J2000) with an uncertainty of 0.42\asec\ (95\% C.L.), using the formula in \citet{Hong2005}. The \chandra\ position is offset from the reported  \swift/XRT and UVOT positions by $\sim9$\asec\ and $\sim3$\asec, respectively.

%%%%%%%%%%%%%%%%%%%%%%%%%%%%

\section{Spectral analysis} \label{sec:spectra}

In this section, we present our spectral analysis of the two X-ray transients from the \nustar\ observations. After describing our \nustar\ data reduction in \S\ref{sec:data_reduction}, We introduce various spectral models in \S\ref{subsec:models} and present spectral fitting to 3--79~keV \nustar\ spectra in \S\ref{subsec:spectral_fit}.  

\subsection{Data reduction}
\label{sec:data_reduction} 

For both transients, we extracted \nustar\ source spectra from a 30\asec\  circular region.  We generated response matrices and ancillary response files using {\tt nuproducts}. We generated background spectra for each transient differently as we describe below. All spectra were grouped with a minimum of 30 counts per bin and fitted using XSPEC (v12.9.1). For both transients, as described below, the source spectra dominate the background over the entire 3-79 keV \nustar\ energy band. 

{\bf \textit{T15:}}
{  We extracted \nustar\ background spectra for T15 from an earlier \nustar\ observation, dated 2014 July 4th (ObsID: 30001002010), which preceded the 2016 outbursts of T15 and T37. These background spectra may be subject to contamination from the nearby NS-LMXB \axj. While T15 and \axj\ were well resolved by \nustar\ as shown in Figure~\ref{fig:nustar_images}, we estimated the level of contamination from \axj\ in the following manner: First, we extracted \nustar\ spectra of \axj\ from the 2014 July and 2016 February \nustar\ observations. By fitting the \nustar\ spectra of \axj, we characterized their spectral shapes and measured the flux variation between the two \nustar\ observations. Then, using the \nustar\ PSF file, we computed the fraction of X-ray photons from \axj\ within the source extraction circle around T15. We scaled the extracted \nustar\  background spectra by taking into account the \nustar\ flux variation and PSF fraction to reflect the time variability of \axj. As a result, we found that the scaled background spectra were negligible ($<2$\%) compared to the T15 source spectra. }

{\bf \textit{T37:}}   To avoid contamination from both T15 and \axj, we extracted background spectra from the 2016 February \nustar\ observation, preceding the onset of the T37 outburst. {  The extracted background spectra were  scaled to reflect the change in the T15 flux (as shown in the left panel of Figure~\ref{fig:swift_lc}); the scaling was determined by comparing the \swift/XRT observations simultaneous with the two \nustar\ observations.  We utilized \swift/XRT data since T15 and \axj\ are not visible in the \nustar\ image (in the right panel of  Figure~\ref{fig:nustar_images}), because the brightness of T37 dominated over other X-ray sources. We found that the contamination level from T15 and \axj\ contributed less than $3$\% of the T37 source spectra. }

\subsection{Spectral models} \label{subsec:models}

Before we present our spectral fitting results in \S\ref{subsec:spectral_fit}, below we describe our spectral models for clarity. All the model components we used for spectral fitting are available in XSPEC. 

\subsubsection{Photo-absorption and dust scattering} 

Photo-absorption and dust scattering in the high density environment around the GC can affect X-ray source spectra significantly. Neutral hydrogen absorption was fitted with the {\verb=tbabs=} model using the abundances of \citet{Wilms2000}. To account for 
the effects of dust scattering, we applied 
a spectral model developed by \citet{Jin2017}. This multiplicative model (hereafter {\verb=dust=}), with parameters such as grain sizes and types, column densities and distances of dust layers, was uniquely determined by fitting the \chandra\ radial profiles of T15. The model requires a foreground dust layer in the spiral arms a few kpc away from the GC with $N_{\rm H} \sim 1.7\times10^{23}$~cm$^{-2}$ \citep{Jin2018}. The column density is consistent with another independent study based on T15 \citep{Corrales2017}. The {\verb=dust=} model was then constructed for each of the two transients and each source extraction region. All spectral models described below are multiplied by the model components {\verb=tbabs=} and {\verb=dust=}.  

\subsubsection{Phenomenological models} 

In order to characterize the overall spectral shapes,  measure X-ray fluxes, and assess the presence of (broad) Fe emission lines for X-ray transients, we first fit  phenomenological models composed of power-law, blackbody, thermal disk, and gaussian line components. X-ray transient spectra are usually characterized by thermal ($kT \la 1$ keV) and non-thermal continuum components accompanied by broad Fe emission lines or absorption features at $E = $~5--10 keV. {\verb=diskbb=} represents multi-temperature thermal emission from the accretion disk, while the blackbody model is used for thermal emission from a NS surface or boundary layer \citep{Lin2007}. Fitting a gaussian line determines the Fe line centroid, equivalent width and line profile. The photon index from a power-law model fit can help ascertain whether the transient is in the low/hard, high/soft, or intermediate state. Our baseline phenomenological models are  {\verb=diskbb+powerlaw+gaussian=} for a BH transient, while a blackbody {\verb=bbodyrad=} component is added for the NS transient case. We also replaced  {\verb=gaussian=} by the {\verb=diskline=} or {\verb=kerrdisk=} model to characterize the Fe line features, as they are capable of modelling an (asymmetric) line profile from a relativistic accretion disk \citep{Fabian1989}. 

\subsubsection{Multi-component spectral models with thermal disk, Comptonization and X-ray reflection} 

Three common features are usually observed in X-ray transient spectra, whether they contain a NS or BH, in nearly all outburst states: (1) thermal emission from an accretion disk, (2) Comptonization by a hot corona, and (3) X-ray reflection off the disk. Thermal emission from the accretion disk is modeled with a superposition of  multi-temperature blackbody emissions,  
with the temperature increasing towards the inner edge of the disk \citep{Shakura1973}. While {\verb=diskbb=} is commonly used, several spectral models have fully implemented the relativistic effects around a spinning BH, e.g. {\verb=kerrbb=} \citep{Li2005}, {\verb=bhspec=} \citep{Davis2005}, and a combination of both called {\verb=kerrbb2=} \citep{McClintock2006} in XSPEC.

Some of the thermal photons, originating either from the accretion disk or NS surface, may be up-scattered by energetic electrons in a hot corona that forms over the compact object and/or inner regions of the disk, resulting in a 
power-law-like spectrum.  {\verb=nthcomp=} is a widely-used model that depicts the Comptonization in a hot corona of seed photons emitted by the accretion disk \citep{Zycki1999}. 

X-ray photons scattered in the corona can illuminate the disk and be reflected into our line of sight,
resulting in a Compton scattering hump, emission lines and absorption edges. {\verb=reflionx=} self-consistently models X-ray reflection spectra by taking into account the temperature gradient and the ionization states in the accretion disk \citep{Ross2005}. {\verb=reflionx=} produces a model X-ray reflection spectrum averaged over inclination angles and assumes for the illuminating source a power-law spectrum with an exponential cutoff and a folding energy fixed at 300 keV. {\verb=reflionx=} is suitable for modelling X-ray reflection in an accretion disk at a large distance from the compact object, where the relativistic effects are negligible. 
Recently, the {\verb=reflionx_hd=} model was developed for higher density accretion disks ($n > 10^{15}$~cm$^{-3}$), as disk density significantly impacts spectral shape and measured Fe abundance \citep{Garcia2016, Tomsick2018, Jiang2019}. 

\begin{figure*}[ht!]
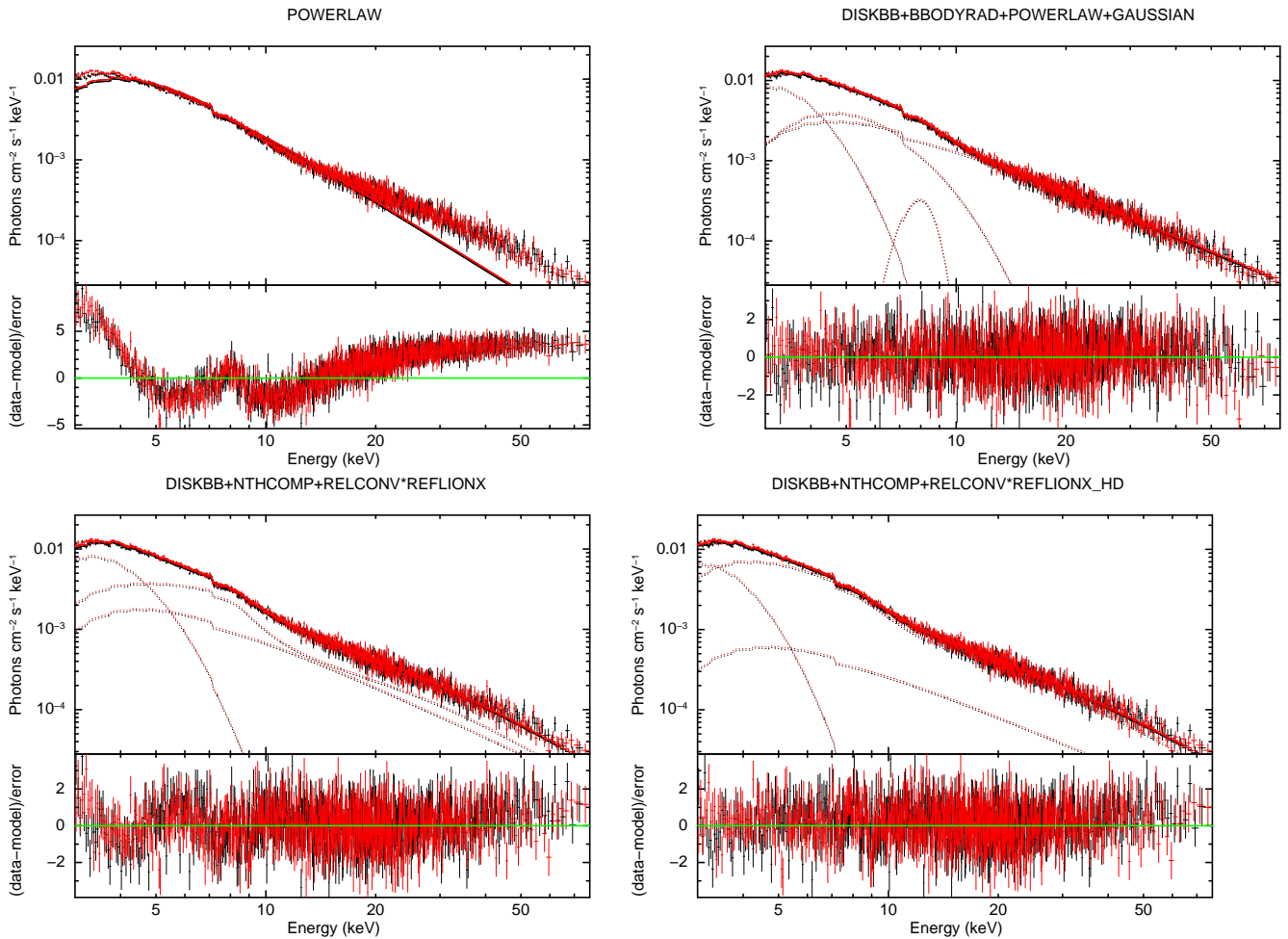

\includegraphics[angle=-90, width=8.5cm]{15_pow_final.eps}
\includegraphics[angle=-90, width=8.5cm]{15_d+bb+pow+g_final.eps}
\includegraphics[angle=-90, width=8.5cm]{15_d+n+rr_final.eps}
\includegraphics[angle=-90, width=8.5cm]{15_d+n+rrhd_final.eps}
\caption{3--79 keV unfolded \nustar\ spectra of T15  fit with a power-law model (top left), {\tt diskbb+bbodyrad+powerlaw+gaussian} (top right), {\tt diskbb+nthcomp+relconv*reflionx} (bottom left) and {\tt diskbb+nthcomp+relconv*reflionx\_hd} model (bottom right). In each panel, the residuals are shown on the bottom.  \nustar\ FPMA and FPMB spectra are shown in black and red, respectively.}
\label{fig:15spectra}
\end{figure*}

In many X-ray transients or XRBs, both the thermal disk emission and reflected X-ray photons are subject to relativistic broadening near the central compact object. Broadened Fe emission lines and absorption edges at $E\approx5$--10~keV are frequently observed in X-ray transient spectra and can be used to constrain 
fundamental parameters, such as BH spin. One can account for the relativistic effects by convolving {\verb=reflionx=} with a broadening function, such as {\verb=relconv=} or {\verb=kdblurr=}, which includes the Kerr metric around a spinning BH \citep{Dauser2010}. The {\verb=relconv=} convolution function has been applied to test whether relativistic broadening is important.

Besides the combination of {\verb=relconv=} and {\verb=reflionx=}, another class of relativistic X-ray reflection models, {\verb=relxill=}, also allows us to measure fundamental parameters such as BH spin and the inner accretion disk radius. {\verb=relxill=} ray-traces all reflected X-ray photons from the disk while taking into account all relativistic  effects  \citep{Garcia2014}. The {\verb=relxill=} model family offers several options for 
illuminating source spectra ({\verb=relxill=} or {\verb=relxillCp=} models for a broken power-law or {\verb=nthcomp=} Comptonization input spectrum, respectively), location ({\verb=relxilllp=} for a lamp post geometry of the corona), or both ({\verb=relxilllpCp=} for a Comptonized illuminating spectrum with the lamp-post geometry).

Here we define several baseline models for fitting the  X-ray transient spectra. First, we fit  both {\verb=diskbb+nthcomp+reflionx=} and {\verb=diskbb+nthcomp+relconv*reflionx=} models to investigate whether the relativistic broadening is required. 
For self-consistency, we link several common parameters between the different model components -- e.g., disk temperature between  {\verb=diskbb=} and {\verb=nthcomp=}; power-law index between {\verb=nthcomp=} and {\verb=reflionx=}. We also use  {\verb=reflionx_hd=} for the high density accretion disk when Fe abundance fits to an unreasonably high value above $A_{\rm Fe}\sim5$. 

{  In contrast to the {\verb=relxill=} models,  {\verb=relconv*reflionx_hd=} is not fully self-consistent since {\verb=reflionx_hd=} assumes a slab-like optically-thick atmosphere with constant density \citep{Ross2005}. 
In reality, X-ray emission comes from an extended accretion disk, and its density should vary over distance from the central compact object  \citep{Svensson1994}. Instead, our usage of {\verb=reflionx_hd=} assumes that X-rays are reflected from a single layer at certain distance  (which we dub photo-ionization radius $R_{\rm ion}$ hereafter). 
Therefore, we performed a sanity check for self-consistency of the parameters determined from fitting the {\verb=relconv*reflionx_hd=} model in the following way. 

In {\verb=reflionx_hd=}, the ionization parameter is defined as $\xi \equiv\frac{4\pi F}{n}$, where $F$ is the total illuminating flux and $n$ is the hydrogen number density.  Assuming that an illuminating source (e.g. a hot corona) emits X-ray photons isotropically, $\xi = \frac{L}{n R_{\rm ion}^2}$, where $L$ is the illuminating luminosity. Since $\xi$, $L$ and $n$ are determined from our spectral fitting, one can derive the photo-ionization radius: $R_{\rm ion} = (L/n\xi)^{1/2}$. 
$\verb=reflionx_hd=$ quotes electron density ($n_e$), which is close to $n$,  since the accretion disk is believed to be  predominantly composed of hydrogen. On the other hand, the relativistic broadening function {\verb=relconv=} outputs the inner radii ($R_{\rm in}$) of the accretion disk. Since X-ray reflection should take place in the accretion disk, we impose $R_{\rm ion} \simgt R_{\rm in}$ as a necessary condition for self-consistency of the model, assuming that the illuminating X-ray source (i.e. hot corona) is located above the central compact object. In \S\ref{subsec:15} and \S\ref{subsec:37}, we examine our fitting results by comparing $R_{\rm in}$ and $R_{\rm ion}$. 

}

\subsection{Spectral fitting} \label{subsec:spectral_fit}

In this section, we present our spectral fitting results for the models described above. 
For joint spectral fitting, we applied cross-normalization constants between \nustar\ module A and B. All errors quoted in the following sections correspond to 90\% confidence levels. For the phenomenological models, we used the {\tt error} command in XSPEC.

For the self-consistent models, we found the conventional XSPEC {\tt error} command to be impractical for calculating errors due to the large number and degeneracy of the parameters in these models. We instead used the {\tt chain} command in XSPEC. This command runs a Markov chain Monte Carlo (MCMC) algorithm to compute the posterior probability distributions for each parameter. Errors are calculated by taking the central 90\% of the sorted values of each parameter in the chain(s), the distribution of which matches the posterior probability distribution of each parameter. We chose to use the Goodman--Weare MCMC sampling algorithm \citep{Goodman2010} as it is affine invariant (i.e. its performance is not affected by the degeneracy of the parameters of interest), making it well suited for our purposes. For all models, we first found the best-fitting parameters using the conventional {\tt fit} command, then initialized the algorithm with 40 walkers and specified a chain length of $10,000$. We ignored, or ``burned,'' the first 1,000 steps (3,000 steps for the T37 {\verb=reflionx_hd=} model) in order to avoid biasing the distribution with parameter values calculated before the chain reached a steady state. We repeated the chain calculation five times, for a total of 50,000 stored steps. We found that this number of elements was sufficient to reach a Rubin-Gelman criterion $<$ 1.1 for each parameter, implying a high confidence in convergence for each parameter \citep{Verde2003}. For a more in-depth discussion of MCMC analysis in X-ray spectroscopy, see, e.g., \citet{Reynolds2012} or \citet{Steiner2012}.
%%%%%%%%%%%%%%%%%%%%%

\subsection{T15} \label{subsec:15}

We fit the 3--79 keV \nustar\ FPMA and FPMB spectra. We froze the hydrogen column density to the value 
($N_{\rm H} =  17 \times10^{22}$~cm$^{-2}$), which was determined by fitting the dust scattering halo profile of T15 \citep{Corrales2017}. 
See Figure~\ref{fig:15spectra} for the \nustar\ spectra of T15 fit with the four models described below and Table \ref{tab:specfit_15} for the fitting results.   In general, we found that the flux normalization difference between the FPMA and FPMB spectra, as calculated by the  {\verb=const=} model, was about 5\%\footnote{Within the typical range for \nustar\ spectra (\url{https://heasarc.gsfc.nasa.gov/docs/nustar/nustar\_faq.html\#\#coadd\_spectra}).}.  
An absorbed power-law or {\verb=diskbb+powerlaw=} model fit results in large $\chi^2_\nu$ values of 5.92 or 2.95, respectively, with significant residuals throughout the spectra (e.g., the top-left panel of Figure \ref{fig:15spectra}). Adding a blackbody component with the best-fit $kT = 1.3$~keV greatly improves the spectral fit ($\chi^2_\nu = 1.1$ for 1326 dof), but a broad emission-like feature centered at $E\sim$8~keV still remains. A  {\verb=gaussian=} line component fits the 8~keV residuals well at the centroid energy $E= 7.90\pm0.07$~keV, with $\sigma= 0.76^{+0.17}_{-0.13}$~[keV] and equivalent width (EW) = $0.22$~keV (see upper right panel in Figure~\ref{fig:15spectra}). The emission feature at $E\sim8$~keV appears to be an artifact of fitting a single gaussian component to the complex Fe features since its centroid is higher than the typical Fe line energies at 6.4--6.9~keV. The  
{\verb=diskbb+bbodyrad+powerlaw+gauss=} model yields an excellent overall fit with $\chi^2_\nu = 0.98$ (1322 dof; Table \ref{tab:specfit_15}). 
Both the thermal disk and blackbody components are required to fit the low-energy spectra at $E \simlt 5$~keV. The best-fit blackbody radius  $3.2\pm0.2$~km at the GC distance (8~kpc) is smaller than the canonical NS radius ($\sim10$~km), but the blackbody radiation may be emitted from hot spots on a NS surface or boundary layer \citep{Lin2007}. {  Otherwise, we note that \citet{Tomsick2018} found a blackbody component with higher temperature (0.7 keV) than the inner disk temperature (0.3--0.4~keV) in the \nustar\ spectra of BH-HMXB Cygnus X-1 during the intermediate state. }

\begin{deluxetable*}{lcccc}[b]
%\tabletypesize{\small}

\tablecaption{Spectral fitting results of T15}
\tablecolumns{4}
%\tablewidth{0.96\linewidth}

\tablehead{ \colhead{Parameter}   &  \colhead{{ powerlaw+diskbb+bbodyrad+gauss}}  &  \colhead{nthcomp+diskbb+relconv*reflionx} &  \colhead{nthcomp+diskbb+relconv*reflionx\_hd\tablenotemark{a}}}
\startdata
$N_{\rm H}$ [$10^{22}$~cm$^{-2}$]  & $17$ (frozen)   & $17$ (frozen) & $17$ (frozen)      \\
$\Gamma$    & $1.87\pm0.03$         & $2.00^{+0.01}_{-0.02}$        & $1.85_{-0.02}^{+0.03}$          \\
$kT_{\rm in}$ [keV] & $0.53\pm0.02$ & $0.681\pm0.002$ & $0.53\pm0.01$ \\
$kT_{\rm bbodyrad}$ [keV] & $1.22\pm0.04$ & - & - \\
$N_{\rm pl}$ or $N_{\rm nthcomp}$ & $ 0.11\pm0.01 $ & $0.037\pm0.008$  & $0.014^{+0.002}_{-0.004}$  \\
$N_{\rm diskbb}$ & $(5.4^{+0.9}_{-0.7})\times10^3$ & $(6.2\pm0.1)\times10^2$ & $(2.8^{+0.5}_{-0.4})\times10^3$ \\
$N_{\rm bbodyrad}$ & $14.7^{+2.3}_{-1.6}$ & - & - \\
$E_{\rm line}$ [keV] & $7.9\pm0.1$ & - & - \\ 
$\sigma_{\rm line}$ [keV] & $0.7\pm0.2$ & - & - \\ 
EW$_{\rm line}$ [keV] & $0.21$ & - & - \\ 
$R_{\rm in}$ [$R_{\rm ISCO}$] & -  &   $1.2^{+0.1}_{-0.2}$      & $1.3^{+0.4}_{-0.2}$       \\
BH spin $a_*$ &  -  &   $>0.985$          & $0.94^{+0.03}_{-0.10}$        \\
Inclination angle [$^\circ$] &  -   &   $65.7^{+0.5}_{-1.3}$        & $64.2^{+0.9}_{-1.6}$      \\
Ionization parameter log($\xi$) & - & $3.77^{+0.04}_{-0.02}$ & $3.32\pm0.02$ \\ 
Electron density $n_e$ [cm$^{-3}$] & - & - & $(1.0^{+9.0}_{-0.2})\times10^{21}$\tablenotemark{b}\\
$A_{\rm Fe}$ &   -  &    $5.0^{+2.1}_{-0.5}$    &  $1$ (frozen)           \\
$F_{\rm unabs}$ (3--79 keV)& $1.23 \times 10^{-9}$  &   $1.14 \times 10^{-9}$  & $1.17\times10^{-9}$  \\
$\chi^2_{\nu}$ (dof) & 0.98 (1322) & 1.08 (1321) & 1.01 (1321) 
\enddata
\tablecomments{All models were multiplied by {\tt const*tbabs*dust}. All error bars are quoted for 90\%
  confidence level. 
  The given fluxes are in units of \eflux.
 }
\tablenotetext{a}{The {\tt reflionx\_hd} model assumes $A_{\rm Fe} = 1$.}
\tablenotetext{b}{The upper limit of the electron density reached the maximum value ($n_e = 10^{22}$~cm$^{-3}$) allowed in the {\tt reflionx\_hd} model.} 
\label{tab:specfit_15}
\end{deluxetable*}

We replaced the power-law component with the more realistic Comptonization model {\verb=nthcomp=}, and the gaussian line by {\verb=reflionx=} (without relativistic broadening). We retained the {\verb=diskbb=} and/or {\verb=bbodyrad=} models to account for the thermal emission in the low-energy band.  For self-consistency, the power-law photon indices  in {\verb=reflionx=} and {\verb=nthcomp=} are linked. This non-relativistic model ({\verb=diskbb+nthcomp+reflionx=}) resulted in a poor fit ($\chi^2_\nu = 1.80$ for 1324 dof), as well as an extremely high Fe abundance of $A_{\rm{Fe}} = 20\pm 3$. 

We then convolved the reflection component with {\verb=relconv=} to smear out the X-ray spectra, in order to account for the relativistic broadening that occurs around a spinning BH \citep{Garcia2014}. We fixed $N_{\rm H}$ to the value ($1.7\times10^{23}$~cm$^{-2}$) measured from the dust scattering study \citep{Jin2017}.  This  {\verb=diskbb+nthcomp+relconv*reflionx=} model improved the fit significantly, with $\chi^2_\nu = 1.08$ (1321 dof), without the blackbody component required in the phenomenological models.  $A_{\rm{Fe}}$ is well constrained to a very high value at $5.0^{+2.1}_{-0.5}$. Using this model, we found that the apparent 8~keV emission bump is due to two relativistically smeared photo-absorption edges of neutral and highly-ionized Fe at $E\sim7$ and $\sim9$ keV, respectively. The inclination angle is also well constrained to $ i = 65\fdg7_{-1\fdg4}^{+0\fdg5}$. Other X-ray transients with similarly high inclination angles have exhibited smeared Fe absorption edges or lines from accretion disk winds  \citep{Ponti2012, Ponti2016b}. The high inclination angle also accounts for the relatively large contribution of the X-ray reflection component compared to that of the corona, as shown in the lower panels of Figure~\ref{fig:15spectra}. The inner radius of the accretion disk $R_{\rm in}$ is fit to $1.2_{-0.1}^{+0.2} R_{\rm ISCO}$, where $R_{\rm ISCO}$ is the radius of the innermost stable circular orbit (ISCO) for a spinning BH.

{  We then replaced {\verb=reflionx=} with its high density version, {\verb=reflionx_hd=}, which allows the accretion disk electron density ($n_e$) to vary from $10^{15}$ up to $10^{22}$ cm$^{-3}$ while fixing $A_{\rm Fe} = 1$. The best-fit model, with $\chi^2_\nu = 1.01$ (1321 dof), yielded $n_e = (1.0^{+9.0}_{-0.2})\times10^{21}$~cm$^{-3}$ (see Table  \ref{tab:specfit_15} and the bottom right panel of Figure~\ref{fig:37spectra}). The small inner disk radius ($R_{\rm in} = 1.3^{+0.3}_{-0.2} R_{\rm ISCO}$) as well as the high BH spin value ($a_* = 0.94_{-0.10}^{+0.03}$) are required to smear out the Fe absorption edges. Given the best-fit ionization parameter ($\xi = 2.1\times10^3$ [erg\,cm\,s$^{-1}$]) and disk density ($n_e = 1\times10^{21}$~[cm$^{-3}$]), we calculated the photo-ionization radius, defined earlier as $R_{\rm ion} = (L/n\xi)^{1/2}$. Using the bolometric (0.01--200~keV) luminosity of the Comptonization plus reflection model components as the illuminating luminosity ($=1.4\times10^{37}$~erg\,s$^{-1}$), we derived $R_{\rm ion} = 26$~[km]. This is compatible with the inner disk radius of $R_{\rm in} \sim 40$~[km], using the best-fit BH spin value and assuming a BH mass of $10 M_\odot$, given the statistical errors and some unknown parameters (e.g., BH mass). 
We found that higher electron densities up to $10^{22}$~cm$^{-3}$ (i.e. the maximum  value allowed in {\verb=reflionx_hd=}) fit the \nustar\ spectra equally well. However, they lead to $R_{\rm ion}$ values that are much smaller than $R_{\rm in}$, thus we do not consider higher density values plausible. 
In addition, the flux normalization of {\verb=diskbb=} yielded $R_{\rm in} \sim 60$~km, which is also consistent with the inner disk radius.
{ We note that inner-disk radii from {\verb=diskbb=} fits may vary by up to $\sim50$~\%, depending on several uncertainties, e.g. spectral hardening   \citep{Merloni2000}; however, these correction factors are not well-determined from our \nustar\ spectra due to the lack of low energy coverage below 3~keV. }
Hence, we conclude that the model parameters are self-consistent with each other. 
We also fit another fully-relativistic model {\verb=diskbb+relxillCp=} and its variations to the T15 spectra.  
However, the Fe abundance fit to an extremely high value of $A_{\rm Fe} \sim 10$ for all flavors of {\verb=relxill=}, including the high density version {\verb=relxillD=}, which should reduce $A_{\rm Fe}$\footnote{Note that we used a preliminary version of the {\tt relxillD} model which is still under development \citep{Garcia2019}.}. Due to the unphysically high Fe abundance values associated with {\verb=relxill=} fitting, we consider the fit results using the {\verb=relconv*reflionx=} model more viable. 

In summary, we conclude that {\verb=diskbb+nthcomp+relconv*reflionx_hd=} is the most plausible model, for several reasons. The higher accretion disk density of $1\times10^{21}$~[cm$^{-3}$]  fits the T15 spectra better, with a lower reduced $\chi^2$ value (1.01 with 1321 dof), likely because the {\verb=reflionx_hd=} model accounts for the excess free-free emission from the dense disk, leading to fewer residuals in the soft part of the spectrum. This model does not require the high Fe abundance ($A_{\rm Fe} = 5$) that the {\verb=reflionx=} model yielded. The small $R_{\rm in} $ value ($1.3~R_{\rm ISCO}$) indicates that the accretion disk is extended inward, close to the BH. The small inner disk radius is also consistent with the flux normalization of {\verb=diskbb=}, as well as with the photo-ionization radius determined from the best-fit parameters of the {\verb=reflionx_hd=} model. The dominant thermal disk component below $\sim5$~keV, the soft photon index ($\Gamma \approx 2$), and the small inner disk radius suggest that T15 was observed in the soft state. }

\subsection{T37} \label{subsec:37}

\begin{figure*}[ht!]
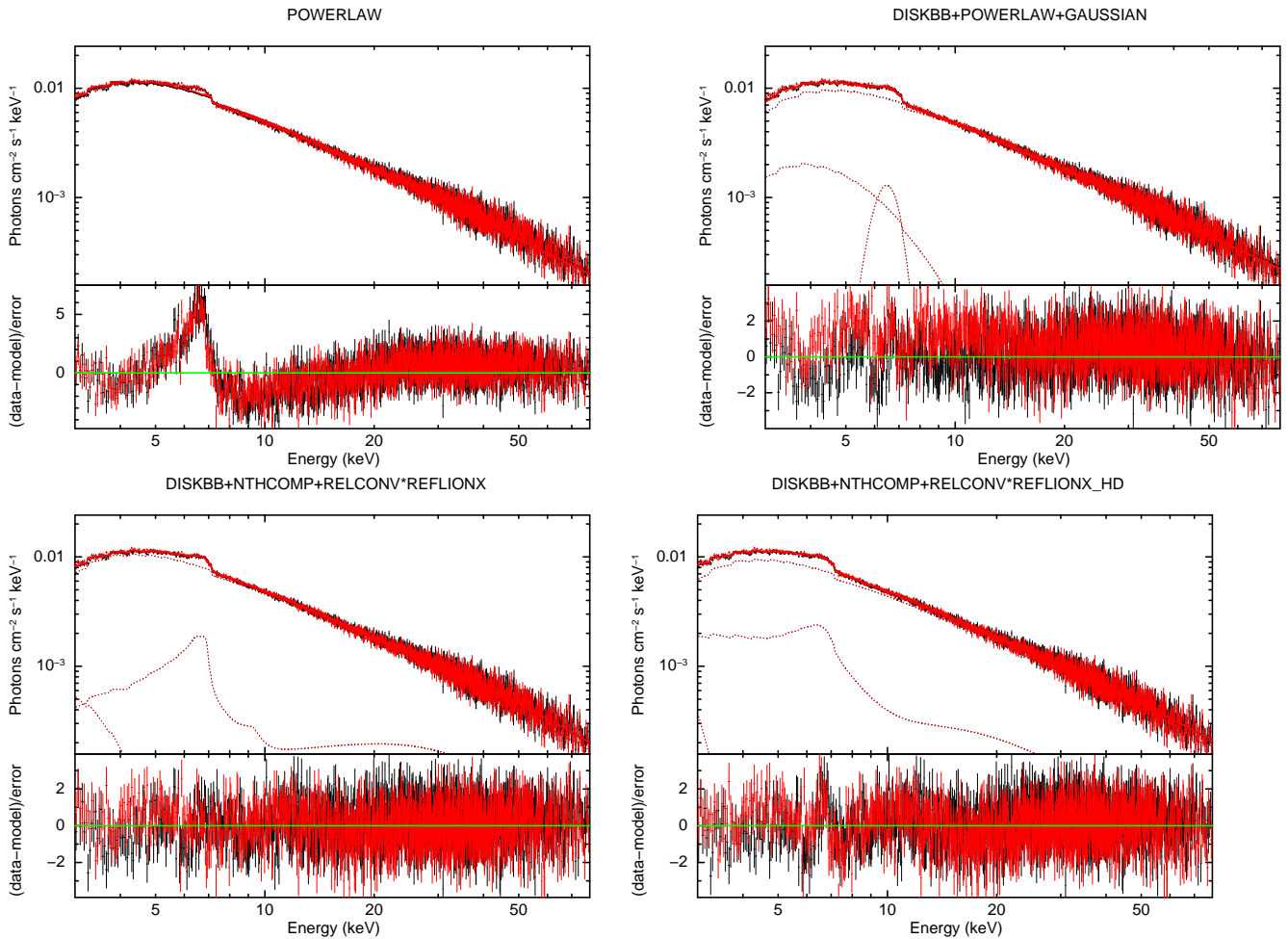

\includegraphics[angle=-90,width=8.5cm]{37_pow_final.eps}
\includegraphics[angle=-90,width=8.5cm]{37_d+p+g_final.eps}
\includegraphics[angle=-90,width=8.5cm]{37_d+n+rr_final.eps}
\includegraphics[angle=-90,width=8.5cm]{37_d+n+rrhd_final.eps}
\caption{3--79 keV unfolded \nustar\  spectra of T37 fit with a power-law model (top left), {\tt diskbb+powerlaw+gaussian} (top right), {\tt diskbb+nthcomp+relconv*reflionx} (bottom left) and {\tt diskbb+nthcomp+relconv*reflionx\_hd} model (bottom right). In each panel, the residuals (data -- model) are shown on the bottom. }
\label{fig:37spectra} 
\end{figure*}

We jointly fit the 3--79 keV \nustar\ FPMA and FPMB spectra. See Figure~\ref{fig:37spectra} for the \nustar\ spectra of T37 fit with the four models described below and Table \ref{tab:specfit_37} for the fitting results. {  The best-fit flux normalization factors were consistent between the FPMA and FPMB spectra within less than 1\%.} An absorbed power-law model yielded $\Gamma$ = 1.6 with $\chi^2_\nu$ of 2.09 (2445 dof). A prominent,  asymmetric Fe emission feature with a red wing, centered around $E=6.5$ keV, was evident in the residuals (see upper left panel in Figure \ref{fig:37spectra}).  Adding {\verb=diskbb=}  and {\verb=gauss=} components  significantly improved the fit ($\chi^2_\nu$ of 1.12 for 2440 dof; see upper right panel in Figure \ref{fig:37spectra}). The best-fit photon index was $\Gamma = 1.51\pm0.005$. The inner disk temperature was  $1.41\pm0.05$~keV. The gaussian model fit to a broad Fe line at $E = 6.45\pm0.02$~keV, $\sigma = 0.51\pm 0.03$ keV and EW = $0.18$~keV. However, some residuals remained around 5--8~keV due to the asymmetric Fe line profile. Replacing the gaussian component with the relativistic emission line {\verb=kerrdisk=} model resulted in a better fit of the Fe line residuals ($\chi^2_\nu  = 1.05 $ for 2438 dof), with the line centroid at $E = 6.90 \pm 0.11$ keV and EW = 0.25~keV.

We proceeded to fit the spectra with our non-relativistic  {\verb=diskbb+nthcomp+reflionx=} model. This model did not fit the data particularly well ($\chi^2_\nu  = 1.22$ for 2439 dof), leaving distinct residuals between 4--7 keV and yielding an Fe abundance of $A_{\rm Fe} =2.0^{+0.5}_{-0.3}$.  We smeared out the reflection component by convolving  {\verb=reflionx=} with {\verb=relconv=}. The  {\verb=diskbb+nthcomp+relconv*reflionx=} model yielded an improved fit around the Fe line features, with $\chi^2_\nu  = 1.04$ (2437 dof), as shown in the lower left panel of Figure~\ref{fig:37spectra}. 
The photon index ($\Gamma = 1.66 \pm 0.01$) and inclination angle ($28^{\circ}\pm2^{ \circ}$) were well constrained. Unlike T15, the smaller inclination angle indicates that fitting the prominent emission Fe lines favors a more face-on viewing angle. The inner disk radius ($R_{\rm in} = 3.9 R_{\rm ISCO}$) is larger than those of T15. However, we found this model problematic since the Fe abundance increased to $A_{\rm Fe}=6.7\pm1.3$.

{  Following the T15 spectral analysis, we replaced {\verb=reflionx=} by the {\verb=reflionx_hd=} model.  Fitting with the  high density reflection model yielded the best-fit disk density of $n_e = (7.2^{+0.5}_{-0.6})\times10^{20}$ cm$^{-3}$. The fit quality, with $\chi^2_\nu$ of 1.10 (2437 dof), was slightly poorer than with the low-density {\verb=reflionx=} version ($\chi^2_\nu  = 1.04$), as shown in the bottom right panel of Figure  \ref{fig:37spectra}. There are several noticeable changes associated with the high-density reflection model fit. (1) Similar to the application of the high density reflection models to BH binary GX~339$-$4 \citep{Jiang2019}, we found that the contribution of the thermal disk component was greatly suppressed, likely because of the enhanced free-free emission in the low energy band, which is a consequence of the high electron density. As a result of the negligible contribution of the {\verb=diskbb=} component, its flux normalization was not well constrained; thus, we obtained only an upper limit. (2) 
We found that the best-fit BH spin and inner disk radius are: $a_* = 0.92^{+0.05}_{-0.07}$ and $R_{\rm in} = 4.1^{+0.8}_{-1.0}~ R_{\rm ISCO}$, respectively. These values were determined mostly from fitting the broad Fe lines and edges at $E\sim$5--10~keV.  The error bars are purely statistical and calculated by the MCMC algorithm described in \S\ref{subsec:spectral_fit}.  (3) As a consistency check, we derived the photo-ionization radius from the best-fit $\xi$, $n_e$ and the bolometric luminosity (0.01-200 keV) for the Comptonization plus reflection model components as the illuminating source ($L = 5\times10^{37}$~erg\,s$^{-1}$). We found $R_{\rm ion} = (\frac{L}{n_e \xi})^{1/2} \approx 100$ [km]. This is comparable to the inner disk radius ($R_{\rm in} = 4.1 R_{\rm ISCO} = 120$~[km], using the best-fit BH spin value and assuming $10 M_{\odot}$ BH mass) obtained from {\verb=relconv=}.  (4) Small residuals are still present at $E \sim 6.6$~keV. We attribute them to the fact that the {\verb=reflionx_hd=} model uses $A_{\rm Fe}$ fixed to 1. 
 The residuals may indicate that Fe abundance is higher than $A_{\rm Fe} = 1$.  Based on the size of the residuals and relative contribution of the reflection model, we estimate that increasing $A_{\rm Fe}$ by $\sim20$\% would fit the residuals if the {\verb=reflionx_hd=} model implements Fe abundance variations in the future. }  
 { Alternately, these Fe line residuals, which manifest in a comparatively narrow line, may be the result of additional disk reflection not modeled by the highly ionized {\verb=reflionx_hd=} component. This would be consistent with irradiation of the less-ionized, more distant outer part of the disk, as modeled for Cygnus X-1 by \cite{Tomsick2018}.}

We then attempted to fit the spectra with the {\verb=relxill=} reflection models.  With the default density of $10^{15}$ cm$^{-3}$, the best fit Fe abundance for the {\verb=diskbb+relxillCp=} model  reached the maximum value of $A_{\rm Fe} = 10$. Similar to the T15 spectra, increasing the disk density to $\approx 10^{19}$~cm$^{-3}$ did not reduce the Fe abundance. Fixing $A_{\rm Fe}$ to a value between 1 and 3 led to a poor spectral fit with $\chi^2_\nu > 1.3$. It is unclear why the high density {\verb=relxill=} model does not reduce $A_{\rm Fe}$ as {\verb=reflionx_hd=} did. 
Investigating the discrepancy between the {\verb=reflionx=} and {\verb=relxill=} models is beyond the scope of this paper, and analyzing other BH transients with some known parameters (e.g., BH mass) is more appropriate for such a comparative study.

{  In summary, we consider the  {\verb=diskbb+nthcomp+relconv*reflionx_hd=} model to  yield  the most plausible results, since the higher accretion disk density obliviates the need of the extremely high Fe abundance measured by the {\verb=reflionx=} model. This model fit results in a high BH spin of $a_* =  0.92_{-0.07}^{+0.05}$. In contrast to T15, the large $R_{\rm in}$ value is consistent with T37 being in the low/hard state (during which the inner edge of the accretion disk is usually located at a large distance from the central BH) when the \nustar\ observation was performed near the peak of the X-ray lightcurve shown in the right panel of Figure~\ref{fig:swift_lc}.} 

\begin{deluxetable*}{lcccc}[b]

\tablecaption{Spectral fitting results of T37}
\tablecolumns{4}
%\tablewidth{0.96\linewidth}

\tablehead{ \colhead{Parameter}   &  \colhead{diskbb+powerlaw+gauss}  & 
\colhead{diskbb+nthcomp+relconv*reflionx}  & 
\colhead{diskbb+nthcomp+relconv*reflionx\_hd\tablenotemark{a}}  }
\startdata
$N_{\rm H}$ [$10^{22}$ cm$^{-2}$]  & $12.3\pm0.4$ & $11.9^{+0.5}_{-0.6}$  & $12.0\pm0.5$   \\
$\Gamma$ & $1.510\pm 0.005$    & $1.59\pm0.01$   & $1.592^{+0.002}_{-0.013}$  \\
$kT_{\rm in}$ [keV] & $1.41\pm 0.05$     & $0.46^{+0.15}_{-0.04}$   &   $0.20^{+0.02}_{-0.01}$  \\
$N_{\rm diskbb}$ & $2.7^{+0.9}_{-0.6}$  & $400^{+700}_{-200}$   & $< 5.2\times10^6$\tablenotemark{b}      \\
$N_{\rm pl}$ or $N_{\rm nthcomp}$ & $0.17\pm0.004$ & $0.17^{+0.03}_{-0.01}$  &  $0.181^{+0.002}_{-0.003}$  \\
$E_{\rm line}$ [keV] & $6.45\pm 0.02$ & - & - \\ 
$\sigma_{\rm line}$ [keV] & $0.51\pm0.03$ & - & - \\ 
EW$_{\rm line}$ [keV] & 0.18 & - & - \\ 
$R_{\rm in}$ ($R_{\rm ISCO}$) & - &   $3.9^{+1.0}_{-0.9}$   &   $4.1^{+1.0}_{-0.8} $   \\
BH spin $a_*$ &  -    &  $0.98^{+0.02}_{-0.04}$    &   $0.92^{+0.05}_{-0.07}$    \\
Inclination angle [$^\circ$] &  -   &    $28\pm2$   &   $21^{+2}_{-3}$   \\
Ionization parameter log($\xi$) & - & $2.71^{+0.02}_{-0.01}$ & $2.76\pm0.06$  \\
Electron density $n_e$ [cm$^{-3}$] & - & - & $(7.2_{-0.6}^{+0.5})\times10^{20}$ \\ 
$A_{\rm Fe}$ &  -  &    $6.7\pm1.3$  &   $1$ (frozen)         \\
$F_{\rm unabs}$ (3--79 keV) & $3.8 \times 10^{-9}$   & $3.69 \times 10^{-9}$ &   $4.1 \times 10^{-9}$      \\
$\chi^2_{\nu}$ (dof)    & 1.09 (2439)   & 1.04 (2437)  & 1.10 (2437)               
\enddata
\tablecomments{All models were multiplied by {\tt const*tbabs*dust}. All errors are quoted for 90\% confidence level. 
  The given fluxes are in units of erg~cm$^{-2}$~s$^{-1}$.
 }
\tablenotetext{a}{The {\tt reflionx\_hd} model assumes $A_{\rm Fe} = 1$.}
\tablenotetext{b}{We obtained an upper limit on the {\tt diskbb} model component whose contribution is negligible in the \nustar\ energy band as shown in the bottom right panel of Figure~\ref{fig:37spectra}. }
\label{tab:specfit_37}
\end{deluxetable*}

\section{Timing analysis} \label{sec:timing}

\begin{figure*}
\plotone{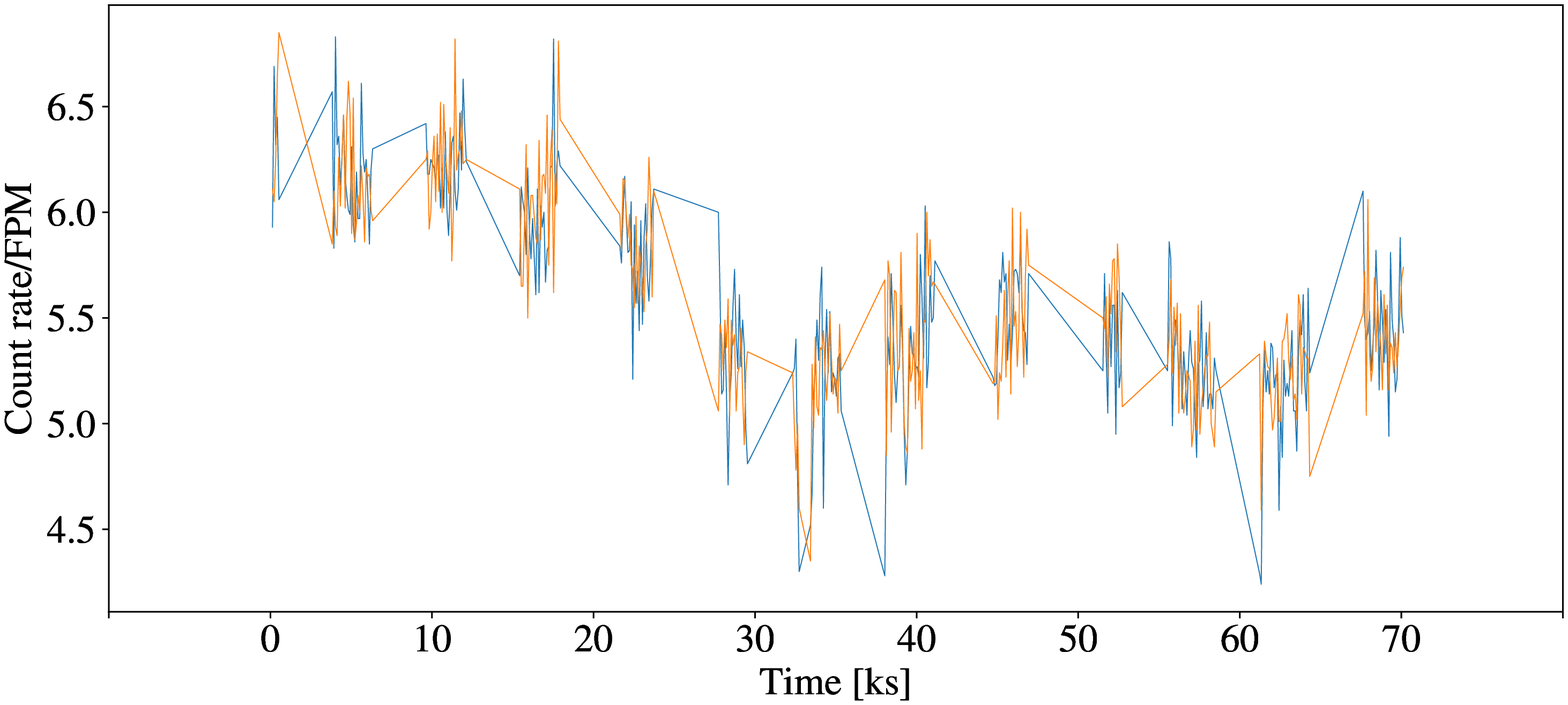} 
\plotone{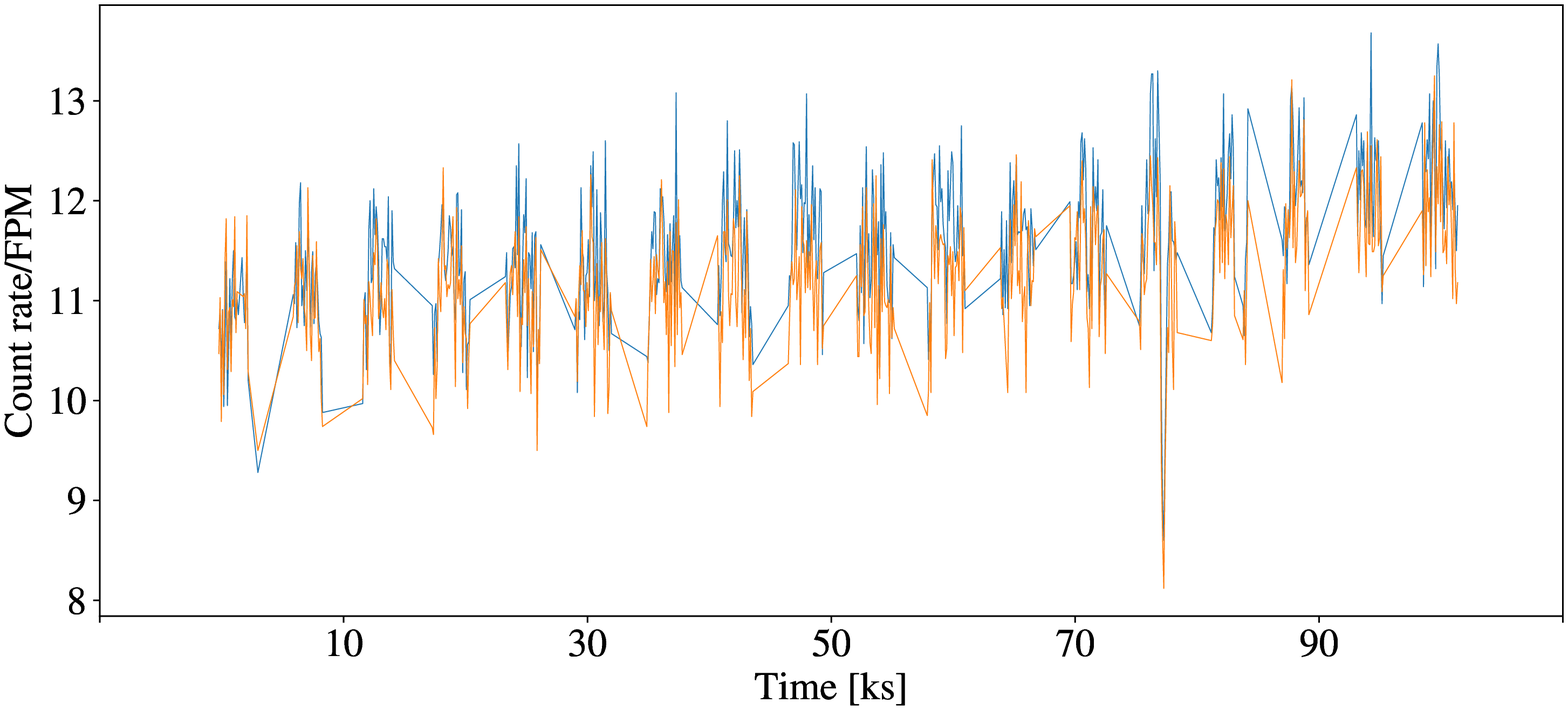} 
\caption{3--79~keV \nustar\ lightcurves of T15 (top) and T37 (bottom) during the \nustar\ observations on 2016 Feb 11 and  2016 May 28, respectively. We binned the lightcurves by 100 sec. In each figure, we show lightcurves in unit of count\,s$^{-1}$ for module A (blue) and B (orange), respectively. {  A dip-like feature at $T \sim 7.8\times10^4$ [sec] was determined to be an instrumental artifact as it was also observed in the background lightcurve.} }
\label{fig:lightcurves}
\end{figure*}

We extensively utilized  the novel X-ray timing analysis software \texttt{Stingray} \citep{Stingray2016} and followed the \nustar\ timing analysis of \citet{Bachetti2015} and \citet{Huppenkothen2017} for generating, fitting, and simulating power density spectra and  co-spectra of the two transients. A co-spectrum represents the real part of the cross-spectrum (i.e., the Fourier transform of module A time-series data multiplied by the complex conjugate of the Fourier transform of module B time series data) and can be used to mitigate instrumental effects caused by the detector dead time \citep{Bachetti2015}.  
After applying the barycentric correction to photon event files using the \nustar\ clock file, we extracted source photons from a $r=30$\asec\ circular region around each transient using {\tt extractor} in FTools.  Similar to the spectral analysis, these extraction regions are chosen to maximize the signal-to-noise ratios by reducing the background contamination from the nearby X-ray transient \axj\ (for T15) or T15 (for T37). Figure~\ref{fig:lightcurves} shows \nustar\ 3--79~keV lightcurves of the two transients after binning by 100~sec. The T15 and T37 lightcurves show  $\sim$5\% and $\sim$6\% variability, respectively, during the \nustar\ observations. {  The source variability is not caused by the background, whose contribution is less than 1\% of the total counts extracted from the $r=30$\asec\ circle around the source.} We did not find any type I X-ray bursts in the source lightcurves.

\subsection{Power density spectra and co-spectra}

To produce a power density spectrum (PDS) and its variations, we used \texttt{HENDRICS} library  in the \texttt{Stingray} software package. \texttt{HENDRICS} has been specifically developed for \nustar\ timing analysis \citep{Maltpynt2015} to take into account dead time effects and observation gaps. 
When the deadtime effects are severe at high count rates (usually above $\sim100$~ct\,s$^{-1}$), it produces wave-like 
features in the white noise. Such an artifact due to the deadtime can mimic QPO-like signals. Given that \nustar\ 3--79~keV count rates per module are $\sim6$ ct\,s$^{-1}$ for T15 and $\sim13$~ct\,s$^{-1}$ for T37, we estimate the deadtime effect is only at a few percent level based on the  product of the \nustar\ readout time $\tau_d \sim 2.5$~msec and the count rate \citep{Maltpynt2015}. Nevertheless, in order to search for QPO signals, we generated co-spectra, whose white noise level is zero even if the \nustar\ timing data are affected by the dead time \citep{Maltpynt2015}. 

The final PDS and co-spectrum are the average of PDS and cospectra calculated in 512-s intervals fully contained inside good time intervals (GTIs).
We binned the source lightcurves with a constant bin size $\Delta T = 0.01$ s and generated PDS in different energy bands. 
Following \citet{Bachetti2015}, a safe interval of 200 seconds subtracted from the start and end of each GTI was applied to remove high background contamination due to the SAA radiation belt. 
HENDRICS automatically discards intervals partly or completely outside GTIs to minimize the spurious frequencies that would be produced by data gaps. 
For completeness, we applied different safe intervals and  time bin sizes for generating PDS. We found no significant differences between them. 
We analyzed PDS in the full 3--79 keV band as well as in three divided bands (3--6, 6--10 and 10--79 keV) roughly corresponding to the thermal disk, broad Fe lines and Comptonization components as discussed in \S\ref{sec:spectra}. 
For a pulsation search, we calculated PDS with smaller time bin size using the interbinning method  \citep{Ransom2002}. No pulsation above the 3-$\sigma$ level was found in the PDS down to 10 msec. 

\begin{figure*}
\includegraphics[width=8.5cm]{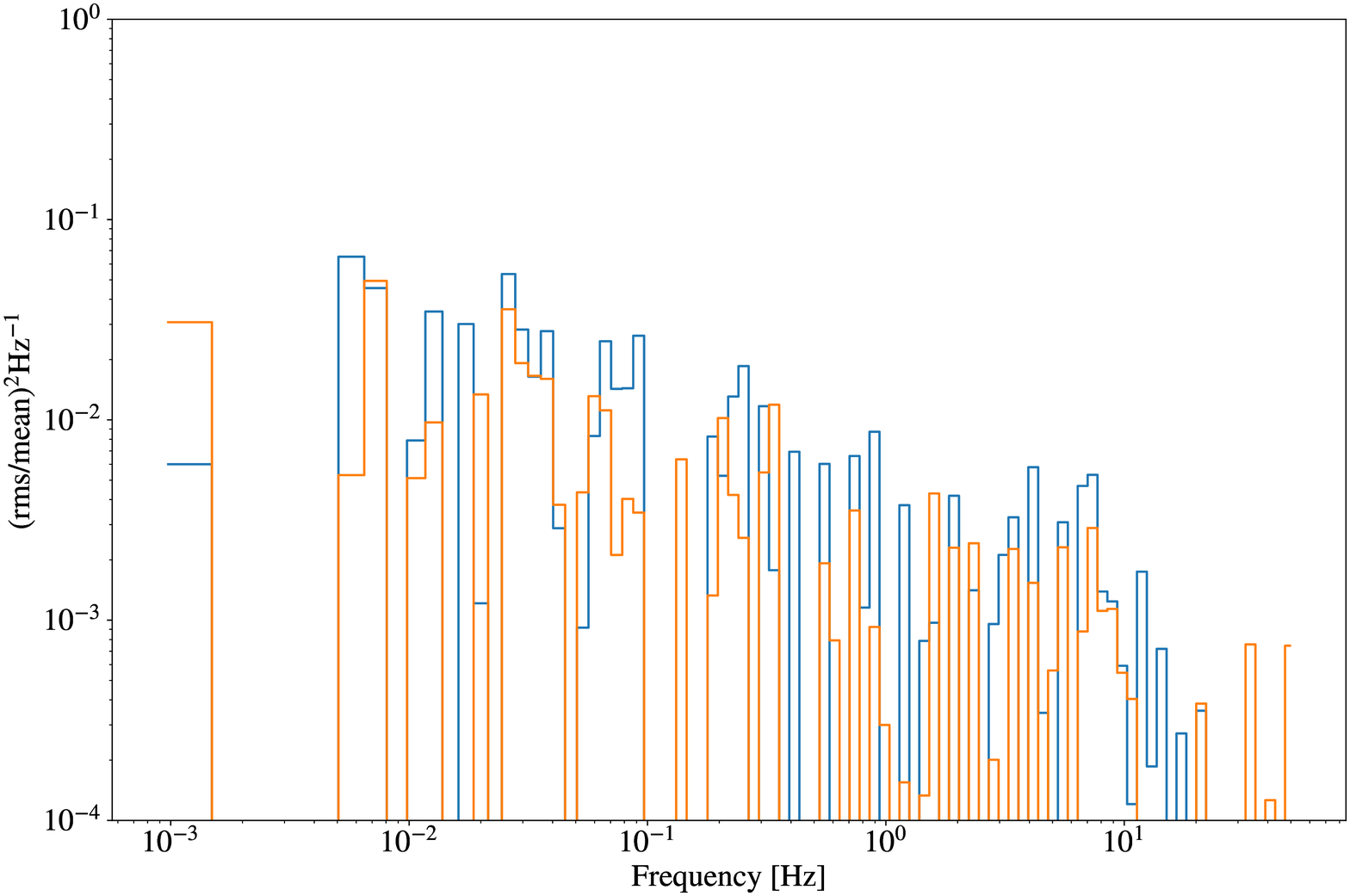}
\includegraphics[width=8.5cm]{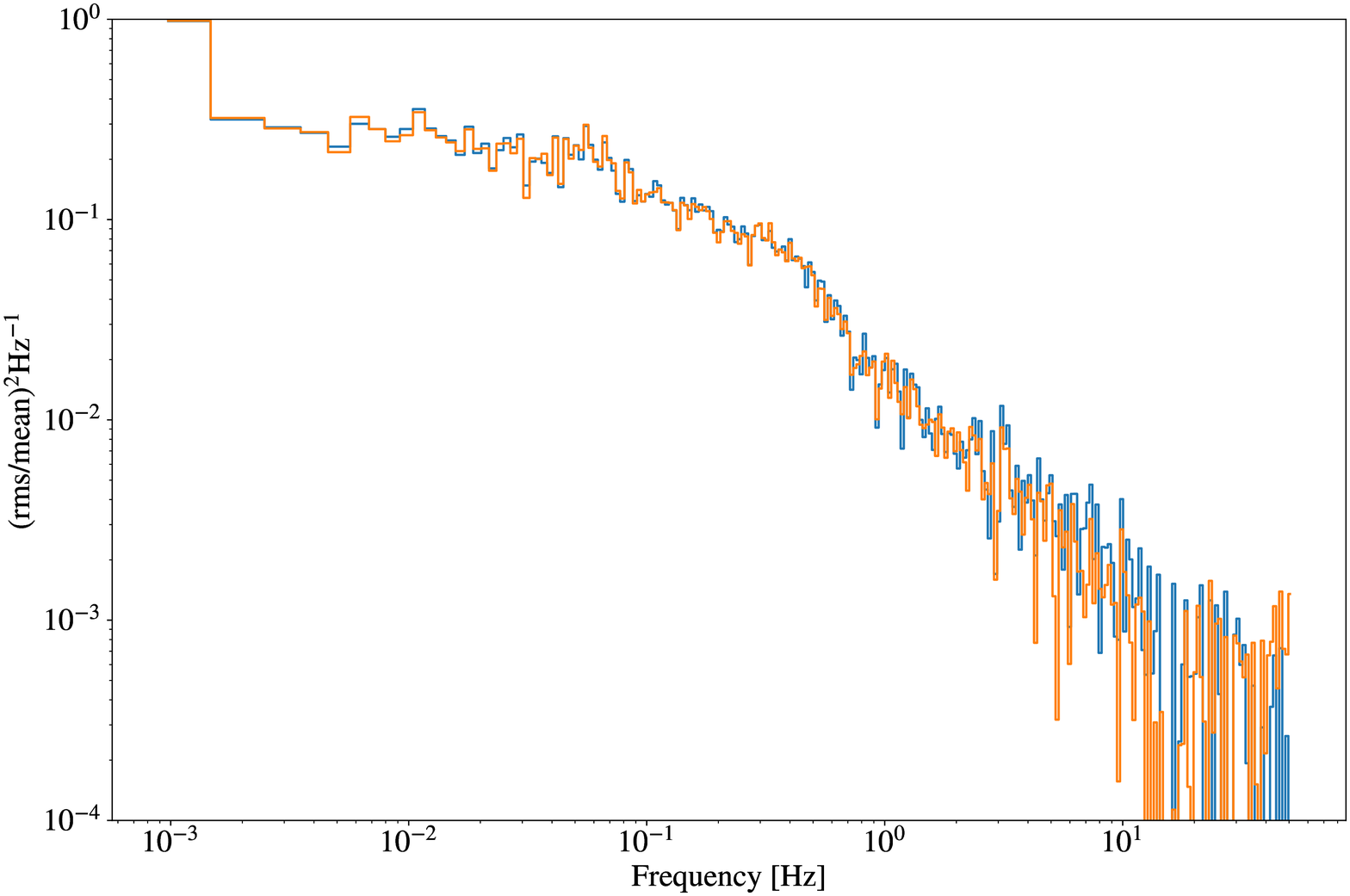}
\caption{\nustar\ PDS of T15 (left) and T37 (right) in the 3--79 keV band. Module A and B PDS are shown in blue and orange, respectively. All PDS were generated using the rms normalization. {  To illustrate the overall shapes better, we rebinned the PDS of T15 and T37 geometrically by a factor of 1.1 and 1.03, respectively.}}
\label{fig:PDS}
\end{figure*}

In Figure~\ref{fig:PDS}, we present \nustar\ 3--79~keV PDS of the two transients in the frequency band $\nu = 0.001$--50~Hz, using the rms normalization. In the plots, we applied geometrical binning to the PDS, by {  a factor of 1.1 (T15) and 1.03 (T37)}, to illustrate the broad-band spectral shapes. Above $\sim20$~Hz, small deviations between the module A and B PDS are seen due to the dead-time effect. Note that the PDS in the sub-divided energy bands (3--6, 6--10 and 10--79~keV) are nearly identical to those in the full energy band. 
Since the white  noise level is subtracted from the PDS in the rms normalization, any positive residuals in PDS represent either red noise or QPOs  due to non-Poissonian time variability, both of which are often observed in X-ray transients. 

It is evident that the T15 PDS are nearly flat with a slight elevation toward the lower frequency, whereas the T37 PDS show a prominent red noise component below $\nu\sim10$~Hz (Figure~\ref{fig:PDS}). While the lack of strong red noise in the T15 PDS is often seen in the intermediate and soft/high state of X-ray transients, the flat top continuum of T37 PDS in the lower frequency band is a common feature in the low/hard  states of BH and NS transients \citep{Vanderklis1995, Belloni2010}. Following \citet{Bachetti2015}, we calculated the fractional rms for T37 as $(32\pm2)$\%, after accounting for deadtime effects.

\begin{figure*}
\includegraphics[width=8.5cm]{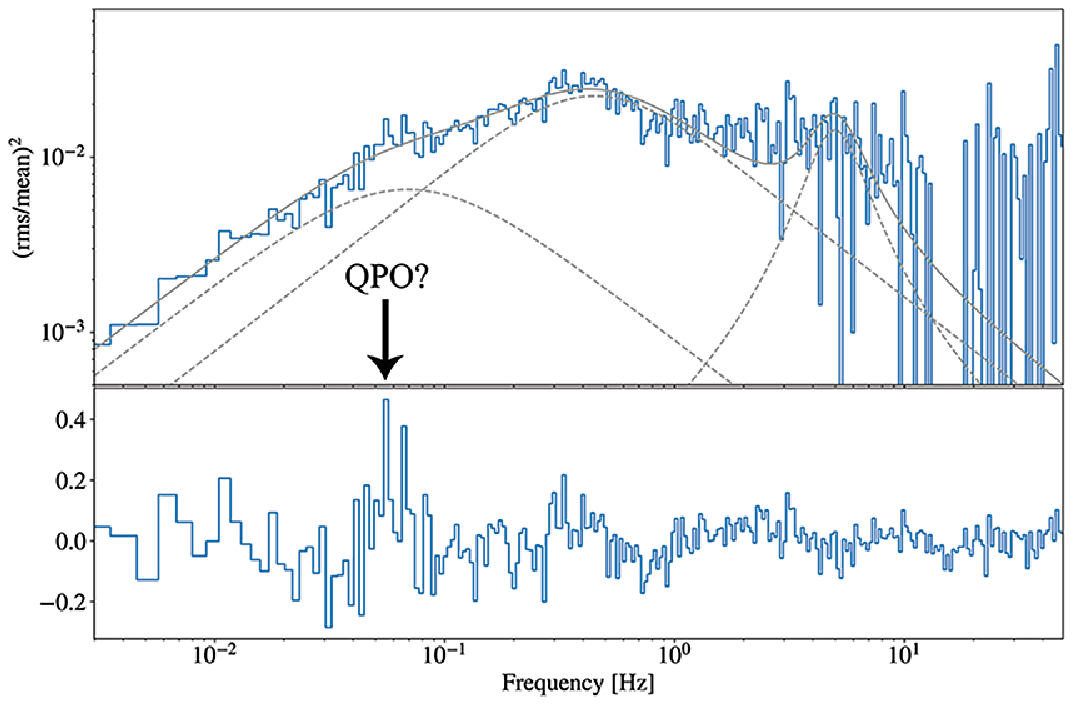}
\includegraphics[width=8.5cm]{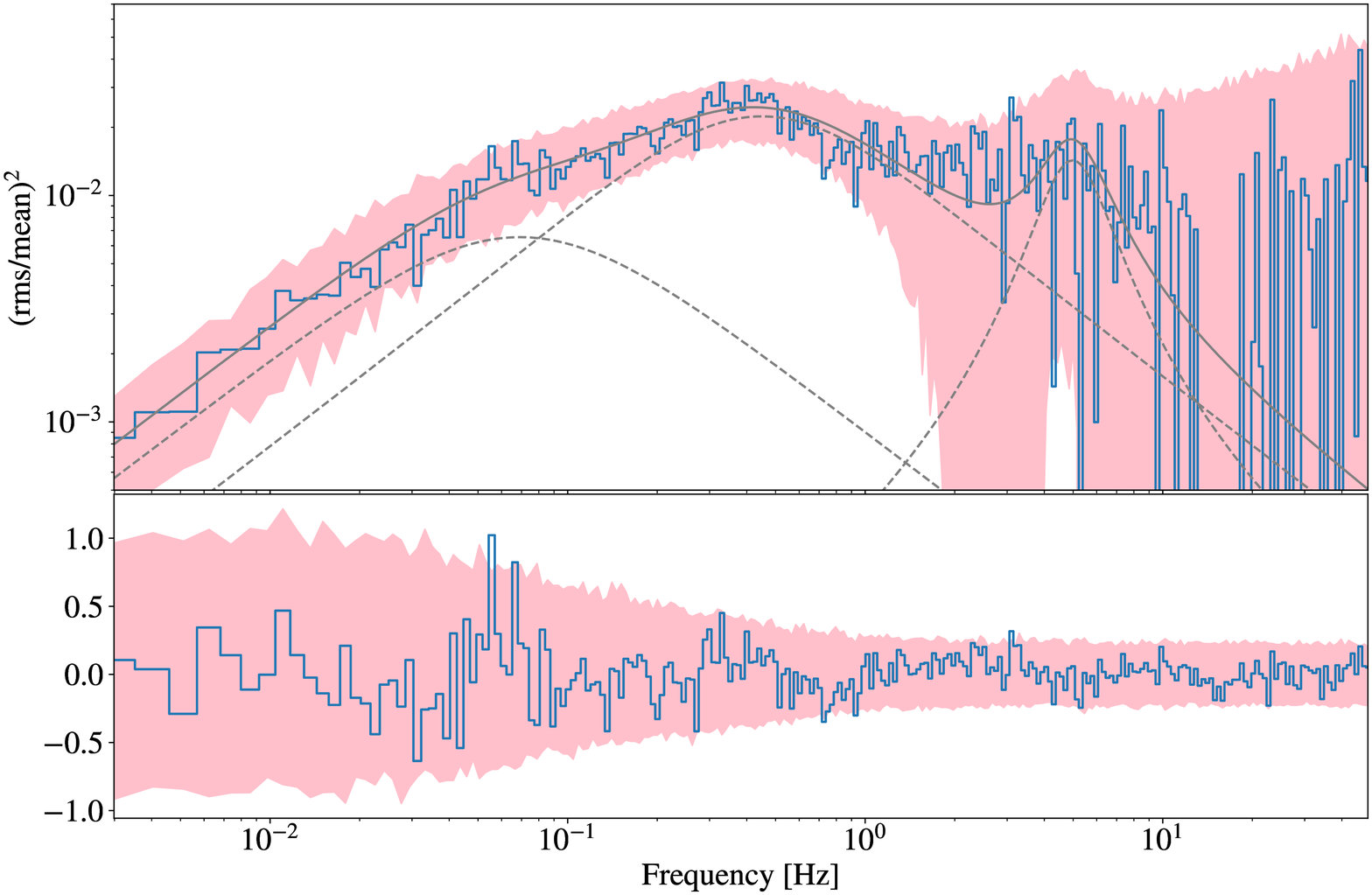}
\caption{\nustar\ 3--79~keV PDS of T37 (left). The PDS is fit to a three Lorentzian model with $\nu_1 = 0.0, \nu_2 = 0.1$ and $\nu_3 = 4.7$~Hz.  The residuals (data--model) are shown on the bottom panel. In the right panel, the same PDS is plotted with the pink contours for 2-$\sigma$  confidence level. The statistical significance of a QPO-like feature at $\sim0.05$~Hz, as indicated by an arrow, is weak at $\sim2$-$\sigma$ level.}
\label{fig:PDS_37}
\end{figure*}

To characterize the T37 PDS better, we first fit its co-spectrum, where any artifacts associated with the deadtime effects are removed, and roughly constrained the model parameters. 
We adopted these model parameters as an initial guess to fit the PDS, then yielded the best-fit parameters using the maximum likelihood method assuming a Gaussian Log Likelihood.  Currently, the proper statistical tests for co-spectra, as presented by \citet{Huppenkothen2018}, have not been implemented in the {\tt  Stingray} software, but given the large number of averaged power spectra, a Gaussian Likelihood is an adequate approximation for the purpose of characterizing the overall PDS. See \citet{Huppenkothen2018} for details. 
We find that the T37 PDS fit well to a model with three Lorentzian functions at $\nu_1 = 0.0$, $\nu_2 = 0.1$ and $\nu_3 = 4.7$~Hz (left panel in Figure \ref{fig:PDS_37}). The presence of multiple Lorentzian components in the PDS is consistent with those of BH transients in the low/hard  state or NS-LMXB atoll sources in the island state \citep{Belloni2010, Vanderklis1995}. The decreasing power above $\sim10$~Hz, as is evident from the highest Lorentzian cutoff frequency at $\sim5$~Hz, is a common feature for BH transients in the low/hard state \citep{Sunyaev2000}. 

In addition, there seems to be an additional QPO-like feature at $\nu\sim50$~mHz. Fitting a Lorentzian model to the line feature yields the line centroid at $\nu_{C} = 52$~mHz with a width $\Delta = 64$~mHz. The quality factor $Q = \nu_C/\Delta \approx 1$ is too small for a typical QPO signal in X-ray transients.  Usually, $Q < 2$ suggests peaked noise \citep{vanderklis2004}. 
Nevertheless, in order to evaluate the significance of the QPO-like signal, we applied the bootstrapping method and empirically determined contour levels of any potential signals in the 0.001--50~Hz band. We repeatedly simulated PDS from the best-fit 3-Lorentzian model and evaluated the likelihood ratio of fitting an additional (4th) Lorentzian line component in the simulated PDS. The right panel of Figure \ref{fig:PDS_37} shows the T37 PDS with 2-$\sigma$ contours for an additional line component. 
Our simulation results yield only weak evidence for the low-frequency QPO at $\nu\sim$50 mHz at $\sim2$-$\sigma$ level. Therefore, we conclude that there is no significant detection of a QPO signal from T37.

\section{{\it Chandra} analysis of the transients during the outbursts and in quiescence} \label{sec:chandra}

In this section, we present \chandra\ analysis of T37 during the 2016 outburst and put constraints on quiescent X-ray fluxes of the two transients.

\subsection{Transient 37 observations}

While \citet{Corrales2017} studied the dust scattering halo from T15 using the \chandra\ observations in February 2016, there has been no publication on the two \chandra\ observations of T37 in July 2016. We investigated the \chandra\ observations of T37 in July 2016 to characterize its spectral state when the outburst flux was declining. 
We analyzed two \chandra\ observations of T37 on 2016 July 12 and 18, 45 and 51 days after the beginning of the X-ray outburst, respectively. These observations occurred over a month after the \nustar\  observations and showed a significant flux decrease compared to the \nustar\ data, and were therefore fit separately. 
We used {\tt dmextract} to extract ACIS source photons from a $r = 3$\asec\ circular region and generated response files using {\tt CIAO 4.9}. Background spectra were extracted from two circular regions with $r = 9$\asec\ and $13$\asec\, respectively.  The extraction radius was adopted to collect X-ray photons scattered from the source position. 

{  First, we fit several models (e.g., absorbed power-law model) to the individual \chandra\ spectra separately and found that the model shape parameters are consistent with each other within statistical errors.  Hence, we fit the two \chandra\ spectra jointly by allowing the flux normalization to vary between them. } Jointly fitting the spectra to an absorbed power-law model yields the best-fit photon index of $\Gamma = 1.67\pm 0.09$ for $\chi^2_\nu = 0.915$ (248 dof).  The 2--8 keV unabsorbed flux is $7.1\times10^{-12}$ and $2.2\times10^{-12}$ \eflux\ for the July 12 and 18 observations, corresponding to $L_X = 5.5\times10^{34}$ and  $1.7\times10^{34}$~erg\,s$^{-1}$, respectively.
An absorbed {\verb=diskbb=} model yields a good fit ($\chi^2_\nu = 0.947$ for 248 dof) and disk temperature $kT = 3.3\pm0.2$ keV. The best-fit disk temperature is too high compared to the typical range of X-ray transients ($kT \simlt 1$~keV), especially in the low/hard state, as evidenced by the photon index and the \swift\ light curve \citep{Remillard2006}. Although an absorbed blackbody model yields $kT = 1.73\pm0.05$ keV for $\chi^2_\nu = 1.03$ (248 dof), the best-fit $N_{\rm H}$ is too low  at $(5.6\pm0.3)\times10^{22}$~cm$^{-2}$.  Therefore, we conclude that the power-law model is most plausible for representing the \chandra\ ACIS spectra.  
  
\subsection{Upper limits on quiescent X-ray luminosity of the 2016 transients}

We attempt to constrain quiescent states of the two transients using the archived \chandra\ observations prior to the X-ray outbursts. 
Neither T15 nor T37 are registered in the \chandra\ source catalogs \citep{Muno2009, Zhu2018}, and no detection has been reported before these outbursts. In order to constrain their quiescent X-ray spectra, we analyzed 45 archived \chandra\ ACIS-I observations of the GC that preceded the X-ray outbursts. We used ACIS Extract (AE) software for spectral analysis  \citep{Broos2010}. We extracted source photons from a region encompassing 90\% of the local point spread function (typically $\sim1$\asec) around the \chandra\ positions of the two transients. Background spectra were extracted from an annular region centered on the source with a background-to-source region area ratio nominally set to 5, avoiding nearby point sources. Response matrices and effective area files were also produced for each observation by AE. More details can be found in \citet{Hailey2018}. 

As a result, we did not detect quiescent X-ray emission of the transients since their 2-8 keV ACIS-I net counts resulted in negative values. Assuming an absorbed power-law spectrum with $N_{\rm H} = 1.7\times10^{23}$ cm$^{-2}$ and $\Gamma = 2.0$ \citep[typical to quiescent BH-LMXB spectra;][]{Plotkin2013}, we obtained 90\% C.L. upper limits on their 2--8 keV fluxes at $2.2\times10^{-15}$ \eflux\ and $2.0\times10^{-15}$ \eflux, corresponding to $L_X < 1.7\times10^{31}$ and $1.6\times10^{31}$~erg\,s$^{-1}$ for T15 and T37, respectively. Note that the upper limit for T15  flux is comparable to $L_X < 5\times10^{31}$~erg\,s$^{-1}$  (0.5--10~keV) obtained by \citet{Ponti2016} who analyzed ACIS-I observation data from 1999 to 2011.

%%%%%%%%%%%

\section{Discussion} \label{sec:discussion}

Below we summarize the results of our analysis of the \nustar, \chandra\ and \swift\ observations of the two transients in 2016. Some of their X-ray spectral and timing properties favor the BH-LMXB  scenario.

\begin{itemize}

\item X-ray spectral models: Broadband 3--79~keV \nustar\ spectra of the two transients are  composed of $kT \simlt  1$~keV thermal disk emission, an X-ray reflection component with relativistically broadened Fe lines, and a power-law like continuum due to Comptonization in hot coronae. These features  are typical of outbursting BH and NS binaries. 
{   We conclude that a combination of {\verb=relconv=} and {\verb=reflionx_hd=} components for modeling X-ray reflection is more plausible given that they yield reasonable fits with lower Fe abundance at $A_{\rm Fe} = 1$. For both transients, we found that the {\verb=relxill=} model and its variations (e.g., {\verb=relxillD=} for a high-density accretion disk case) fit to extremely high Fe abundance values above $A_{\rm Fe} \sim 6$. In the {\verb=relconv*reflionx_hd=} model, we tailored the disk density fit so that the photo-ionization  radius (which was derived from the best-fit ionization parameter and accretion disk density of the {\verb=reflionx_hd=} model) is comparable to or larger than the inner disk radius determined from the relativistic convolution model. It would be useful to perform a systematic study that compares the {\verb=reflionx=} and {\verb=relxill=} models on other BH transients with some known parameters (e.g., BH mass). } 

\item {  BH spin measurements: 

{ These 2016 observations offer the first spin measurements of a BH transient within 100 pc of the GC that utilize broad-band X-ray reflection spectroscopy with \nustar.}
A fast spinning BH (with $a_* \simgt 0.8$) is consistent with the broadened Fe atomic features. The high spin values suggest that the transients contain BHs, since $a_* \sim 0.7$ for a maximally rotating NS (this value was theoretically predicted based on various nuclear equations of state), and is much smaller for observed accreting NS \citep{Cook1994, Miller2015}. For example, $a_* = 0.15$ for the NS-LMXB 4U~1728$-$34, which has a 2.75 [msec] spin period,  assuming that its NS mass is 1.4 $M_\odot$   \citep{Sleator2016}.  Much like some BH transients in the solar neighborhood (e.g., Cyg X-1, 4U 1630$-$472, GRO~J1655$-$40; see \citet{Reynolds2016} and \citet{Middleton2016} for a compilation of previous BH spin measurements of X-ray binaries), the two \swift\ transients in 2016 also show high BH spin values in the range of $a_* \sim0.84 - 0.97$. These values are close to the theoretical upper limits on BH spin due to the radiation effects \citep[$a_* = 0.998$; ][]{Thorne1974} and magnetic fields in accretion disks  \citep[$a_* \sim 0.9$; ][]{Gammie2004, Krolik2005}.  } 

\item Bolometric luminosity: The 3--79~keV luminosity, measured by \nustar, is $8.4\times10^{36}$ and $2.8\times10^{37}$~erg\,s$^{-1}$ for T15 and T37, respectively, well above the luminosity range ($\sim10^{36}$~erg\,s$^{-1}$) of very faint X-ray transients  \citep{King2006}. {  Note that the X-ray luminosity for T37 occurs near the peak of the 2-10 keV \swift\ X-ray lightcurve (the right panel in Figure~\ref{fig:swift_lc}). However, T15 may have been brighter between November 2015 and February 2016,  when the GC was not observable by X-ray telescopes, than during the \nustar\ observation.  Following \citet{Motlagh2019}, we calculated the bolometric luminosities in the 0.01--200~keV band by correcting for the inclination angle effect on the thermal disk component. The bolometric luminosities ($L_{\rm bol}$) are  $2.1\times10^{37}$ and $5.3\times10^{37}$~erg\,s$^{-1}$ for T15 and T37, respectively. Accordingly, their $L_{\rm bol}/L_{\rm Edd}$  ratios are  1.7\% (T15) and 4.2\% (T37), where $L_{\rm Edd}$ is the Eddington luminosity, assuming that they contain a $10M_\odot$ BH. These values are within the range of $L_{\rm bol}/L_{\rm Edd}$ at transition between the hard and soft states \citep{Maccarone2003, Kalemci2013, Motlagh2019}. 
The boundary between the hard and soft states, based solely on $L_{\rm bol}/L_{\rm Edd}$, has lately become more ambiguous - for example, some BH transients remained in the soft state when $L_{\rm bol}/L_{\rm Edd}\sim0.03$\% \citep{Tomsick2014}, which is much lower than the typical range for the soft state. Hence, we used the spectral and timing properties (e.g. power-law photon index and fast variability) to determine the spectral states of the transients, as discussed below. }

\item Quiescent X-ray emission: We found no quiescent \chandra\ counterpart or previous X-ray outbursts at the positions of the two transients. Identification of their IR counterparts is ambiguous, as multiple IR sources  within the \chandra\ error circles did not show time variability during the X-ray outburst of T15 \citep{Ponti2016}. However, the X-ray variability and spectral evolution indicate that the transients are likely LMXBs  \citep{Ponti2016}. Our analysis of the \chandra\ ACIS-I and ACIS-S observations preceding the 2016 outbursts determined that their quiescent X-ray luminosities are $\simlt 2 \times10^{31}$~erg\,s$^{-1}$. {  The faintness of their quiescent states  is more consistent with BH-LMXBs, as \citet{Garcia2001} and \citet{Padilla2014}  found that NS-LMXBs are generally brighter ($L_{\rm X} \simgt 10^{32}$~erg\,s$^{-1}$) than BH-LMXBs in quiescent states, although some of the soft X-ray emission from quiescent NS-LMXBs may be due to thermal emission from the NS surface. } 

\item {  Timing analysis: No pulsations or type I X-ray bursts have been detected from these sources  during the \nustar, \swift\ and \chandra\  observations.  The non-detection of these NS-LMXB signatures supports the BH-LMXB scenario. }

\item T15: The X-ray spectra of T15 are characterized by a soft continuum spectrum with $\Gamma \approx 2$, a more significant thermal disk component than T37, a small inner disk radius of $R_{\rm in} \sim  R_{\rm ISCO}$, and a low variability level in the \nustar\ PDS. These features indicate that T15 was in the soft state during the \nustar\ observation \citep{MunozDarias2011}. This is also supported by the \xmm\ observation \citep{Ponti2016} and the time evolution of T15 as shown in the \swift/XRT light curve (Figure \ref{fig:swift_lc}), which suggests increasing and near-peak flux. The blackbody component (which suggests a NS binary) is not required to fit the \nustar\ + \swift\ spectra with our physically motivated models, and we conclude that its presence in preliminary fitting is an artifact of applying a simple, phenomenological model with a single gaussian component fitted to the complex Fe atomic features at $E \sim $5--10~keV. \citet{Tomsick2018} also found a blackbody component with higher temperature than the inner disk temperature in the \nustar\ spectra of BH-HMXB Cygnus X-1 during the intermediate state. Our 3--79~keV \nustar\ spectral analysis suggests a high inclination angle of $i\approx 65^\circ$. The lack of Fe absorption lines in the \xmm\ data \citep{Ponti2016} or dips in the X-ray lightcurves could be due to the fact that the source is not inclined highly enough ($i \simgt  70^\circ$).  
 
\item T37: T37 was observed by \nustar\ within $\sim2$ weeks after the first detection by \swift/XRT. The hard photon index of $\Gamma \approx 1.6$, in addition to the subdominant or negligible thermal disk component and the larger $R_{\rm in} \simgt 3 R_{\rm ISCO}$, suggests that the source was in the low/hard state during the \nustar\ observation. In contrast to T15, the smaller inclination angle ($i \sim 25^\circ$) results in more prominent Fe emission features in the \nustar\ spectra, as the source is in a more face-on orientation. The T37 PDS fit to three broad Lorentzian profiles with an rms level of $\sim 30$\%, and showed declining variability above $\nu\sim10$~Hz. These features are typically observed in BH transients in the low/hard state \citep{Sunyaev2000}. A potential QPO signal at $\nu \sim50$~mHz was detected at a $\sim2$-$\sigma$ level, and thus is not statistically significant.

\end{itemize}

The follow-up ToO observations of the \swift\ transients in 2016 demonstrated that broad-band X-ray data (with \nustar) and precise source localization (with \chandra) are important for characterizing the spectral states/parameters and inferring the nature of X-ray transients. We will continue observing X-ray transients in the GC through the approved \chandra\ + \nustar\ ToO program in the \chandra\ GO cycle 21, starting in the beginning of 2020.

%%%%%%%%%%%%%%%%%%%%%%%%%%%%%%%%%%%%%%%%%%%%%%%
  
\acknowledgments

This work used data from the \nustar\ mission, a project led by the 
California Institute of Technology, managed by the Jet Propulsion Laboratory, and funded by NASA. We thank  the \nustar\ Operations,
Software and Calibration teams for support with the execution and analysis of these observations. We made use of the \nustar\ Data Analysis Software (NuSTAR-DAS) jointly developed by the ASI Science Data Center (ASDC, Italy) and the California Institute of Technology (USA). 
We acknowledge the scheduling, data processing, and archive teams from the Chandra X-ray Observatory Center, which is operated by the Smithsonian Astrophysical Observatory for and on behalf of the National Aeronautics Space Administration (NASA) under contract NAS8-03060. G.P. acknowledges financial contribution from the agreement ASI-INAF n.2017-14-H.0. C.J. acknowledges the National Natural Science Foundation of China through grant 11873054. D.H. acknowledges support from the Natural Sciences and Engineering Research Council of Canada (NSERC) Discovery Grant and the Canadian Institute for Advanced Research (CIFAR) Azrieli Global Scholars program. 
We would like to thank Daniela Huppenkothen for insightful discussions on \nustar\ timing analysis and {\tt Stingray} software. We acknowledge help from Ninad Nirgudkar and Gabriel Lewis Bridges for generating the \swift/XRT lightcurve plots.

\bibliography{gc_transients}
%\begin{thebibliography}{}

%\end{thebibliography}



\end{document}